\documentclass{article}

\usepackage[nonatbib,final]{neurips_2020}

\usepackage[utf8]{inputenc} 
\usepackage[T1]{fontenc}    
\usepackage{hyperref}       
\usepackage{url}            
\usepackage{booktabs}       
\usepackage{amsfonts}       
\usepackage{nicefrac}       
\usepackage{microtype}      
\usepackage{graphicx,caption} 
\usepackage{amssymb,amsmath} 
\usepackage{cleveref} 
\usepackage[numbers]{natbib} 
\usepackage{tikz}
\usepackage{wrapfig}
\usepackage[page]{appendix}

\usepackage{multirow}
\def\arraybackslash{\let\\\tabularnewline}
\usepackage{booktabs}

\usepackage{enumitem}

\usepackage{standalone} 
\usepackage{longtable}
\usepackage{hyperref}

\usepackage{titlesec}
\titlespacing{\section}{0pt}{\parskip}{-\lineskip}
\titlespacing{\subsection}{0pt}{\parskip}{-\lineskip}
\titlespacing{\subsubsection}{0pt}{\parskip}{-\lineskip}

\newcommand{\e}{{\rm e}}

\newcommand{\R}{{\mathbb R}}

\newcommand{\Ical}{{\mathcal I}} 
 
\newcommand{\Kcal}{{\mathcal K}} 
\newcommand{\Lcal}{{\mathcal L}}

\newcommand{\Tcal}{{\mathcal T}}

\newcommand{\iv}{\sigma}
\newcommand{\ivth}{\sigma_{\theta}}
\newcommand{\wth}{\omega_{\theta}}

\newcommand{\wnn}{\omega_{\rm nn}}
\newcommand{\ivpr}{\sigma_{\rm prior}}
\newcommand{\wpr}{\omega_{\rm prior}}
\newcommand{\lv}{\sigma_{\rm LV}}
\newcommand{\comp}{\mathop{\bigcirc}}
\newcommand*\nr[1]{n^{\rm r}_{#1}}

\newcommand{\datamkt}{\Ical_{\rm 0}}

\newcommand{\datacalbut}{{\Ical_{\rm C45}}}
\newcommand{\dataasy}{{\Ical_{\rm C6}}}
\newcommand{\dataatm}{{\Ical_{\rm atm}}}
\newcommand{\watm}{w_{\rm atm}}

\newtheorem{proposition}{Proposition}[section]

\newtheorem{definition}[proposition]{Definition} 
 
\newtheorem{remark}[proposition]{Remark}

\usepackage{float}

\title{Deep Smoothing of the Implied Volatility Surface}

\author{%
Damien Ackerer \\
UBS\\
Z\"urich, Switzerland \\
\texttt{damien.ackerer@epfl.ch} \\
\AND
Natasa Tagasovska\\
Swiss Data Science Center\\
Lausanne, Switzerland \\
\texttt{natasa.tagasovska@sdsc.ch} \\
\And
Thibault Vatter \\
Department of Statistics \\
Columbia University\\
New York, USA \\
 \texttt{thibault.vatter@columbia.edu} \\
}
\date{First version : June 2019 \\ \vspace{0.25em} This version : June 2020}

\begin{document}

\maketitle

\begin{abstract}
We present a neural network (NN) approach to fit and predict implied volatility surfaces (IVSs).
Atypically to standard NN applications, financial industry practitioners use such models equally to replicate market prices and to value other financial instruments.
In other words, low training losses are as important as generalization capabilities.
Importantly, IVS models need to generate realistic arbitrage-free option prices, meaning that no portfolio can lead to risk-free profits.
We propose an approach guaranteeing the absence of arbitrage opportunities by penalizing the loss using soft constraints.
Furthermore, our method can be combined with standard IVS models in quantitative finance, thus providing a NN-based correction when such models fail at replicating observed market prices.
This lets practitioners use our approach as a plug-in on top of classical methods.
Empirical results show that this approach is particularly useful when only sparse or erroneous data are available.
We also quantify the uncertainty of the model predictions in regions with few or no observations.
We further explore how deeper NNs improve over shallower ones, as well as other properties of the network architecture.
We benchmark our method against standard IVS models.
By evaluating our method on both training sets, and testing sets, namely, we highlight both their capacity to reproduce observed prices and predict new ones.

\end{abstract}


\section{Introduction}

The implied volatility surface (IVS) is a key input for computing margin requirements for brokers, quotes for market makers, prices of exotic derivatives for quants, and strategies positions for traders.
As a result, tiny predictions errors can lead to dramatic financial losses.
But standard IVS models often lack the ability to flexibly reproduce market prices and value other instruments without quotes. 
In this paper, we merge known ideas from different fields, namely machine learning (ML) and mathematical finance, to build a new solution for a non-trivial and relevant financial problem: the interpolation and extrapolation of the IVS.
More specifically, we use a neural network (NN) to correct the IVS produced by any standard model.
This lets practitioners plug our method on top of existing approaches while offering an unprecedented trade-off between flexibility and computational complexity.
This problem was initially brought to us by a financial institution willing to improve its existing solution and a version of this method is being tested in production by a financial institution, where thousands of models are continuously updated throughout the day and used as inputs to various quantitative tools.

Practitioners have a growing interest to leverage NNs as flexible predictors for applications requiring an understanding of the complex dynamics of financial markets. 
And the ML community continues to build better tools and understanding deep models.
But leveraging domain expertise about a specific problem is often difficult, and still an active field of research.
In this paper, we show how knowledge from both ML and mathematical finance can be merged to build a well performing and consistent hybrid model.
Our application focuses on options, yet, the same approach could be similarly applied to other financial problems.

\textbf{Problem setting and background context.}
An \emph{option} is a financial contract giving the option holder the right to buy (a call option) or the right to sell (a put option) an asset, such as a stock or a commodity, for a predetermined price (the \emph{strike price}) on a predetermined date (the \emph{expiry} date).
An initial \emph{premium} must be paid to the option seller in order to acquire today the right to buy or sell and asset in the future at, possibly, a preferential price.
The standard, textbook approach to model option pricing is based on the so-called Black-Scholes (BS) formula.
The Black-Scholes formula provides a closed-form formula for option prices for a specific stock price model, the geometric Brownian motion (GBM).
However, the formula builds on unrealistic assumptions such as continuous price trajectory and trading, absence of market frictions such as bid-ask spread and integer contract size, and normality of log-returns.
Options are traded on financial exchanges, and their prices typically invalidate the GBM model.
The Black-Scholes formula is a convenient one-to-one mapping between a price and a volatility parameter that is preferred by practitioners for a variety of reasons.
For example, it allows them to easily compare the price of options with different contract strike prices and maturities.
As a consequence, a lot of domain expertise has been developed for the so-called \emph{implied volatility surface (IVS)}: the continuous representation of this volatility parameter expressed as a function of the strike price and of the expiry at a given point in time.
However, only a finite number of option prices are observed in practice.
In addition, quoted option prices should not allow \textit{arbitrage opportunities}, that is, constructing a portfolio that may generate profits at a zero initial cost.
Therefore the two main challenges when modeling the implied volatility surface are, first, to ensure that the corresponding option prices do not allow arbitrage opportunities and, second, the generalization to an entire surface given a limited number of observations.
Such a construction allows the IVS to be used as input to price financial derivatives in a consistent way, and that enables effective risk-management.
It is then worth noting that some commonly used models, such as SABR~\cite{hagan2002managing} and SVI~\cite{gatheral2004parsimonious}, may not be arbitrage-free for some parameters values, see for example~\cite[Section~3]{roper2010arbitrage}, which contradicts real-life scenarios.


\textbf{Short literature review.}
As neural networks and machine learning in general, prevail almost all aspects in science and industry, finance applications have also been impacted  \citep{hernandez2007garch, genccay2001pricing, hernandez2013gaussian, cao2003support, heaton2017deep}.
Deep learning have been studied with applications to option pricing in \citep{sirignano2018dgm, han2017overcoming, fujii2017asymptotic, becker2019deep, hutchinson1994nonparametric, liu2019neural, liu2019pricing}.
These papers exploit the well-known universal approximation property of neural networks~\citep{hornik1989multilayer, hornik1991approximation, leshno1993multilayer}.
Several applications of NN models for implied volatility smoothing exist, see~\cite{dugas2001incorporating, zheng2018machine, ludwig2015robust, itkin2015sigmoid, yang2017gated} with more details in Appendix~A.
\cite{zheng2018machine} in his PhD thesis also studied the use of soft constraints to train an arbitrage-free model.
Extensive domain specific knowledge has been built on the IVS.
In this work, we will use the no-arbitrage constraints on the IVS derived in~\cite{roper2010arbitrage}, and later refined in~\cite{guo2016generalized}, the moment condition of~\cite{lee2004moment}, and the surface SVI (SSVI) model of~\cite{gatheral2014arbitrage}.
A more exhaustive literature review is available in Appendix~\ref{sec:litrev}.


\textbf{Summary of contributions.}
We present a new methodology to correct, interpolate, and extrapolate the implied volatility surface in an arbitrage-free way.
We achieve this by modeling the implied total variance as a product of a neural network and a prior model, and by penalizing the loss using soft constraints during training so as to prevent arbitrage opportunities.
The prior model should be a standard valid model for the total variance that will guide the general shape of the predictions, two examples are the Black-Scholes model and the SSVI model of~\cite{gatheral2014arbitrage}.
The neural network is thus acting as a corrector of the prior model, enhancing its ability to reproduce observed market prices.
The soft-constraints specification is guided by theoretical results from mathematical finance.
The two elements together allow to build a realistic, flexible, and parsimonious model for the IVS.
We benchmark our method against standard models and study its performance both on training and testing sets, as well as on synthetic data, and real market prices of contracts on the S\&P~500 index.
Numerical experiments suggest that our method appropriately captures the features of standard option pricing models.
We show that increasing model capacity generally leads to better fits, and that the soft constraints generally help decreasing the fitting error and the convergence speed.
Similar results are obtained when applying our method to real data, where an ablation study shows that constrained learning helps producing better volatility surfaces, both in interpolation and extrapolation.

\textbf{Novelty and significance.}
Since options and related derivatives are actively exchange traded contracts that are used for risk-management and investment purposes, their fair valuation requires a reliable model for the IVS.
The main ambition of this work is bridging the existing gap between a traditional challenge in finance and recent developments from the deep learning community, resulting in a trust-worthy, computationally efficient option pricing framework.
This work is the first peer-reviewed published work to use soft-constraints to guide the training of a deep NN model for the IVS.
Previous work used hardwired constraints which limits tremendously the neural network flexibility~\cite{dugas2001incorporating}, or used constraints on the option price surface which requires additional transformations at each training step~\cite{ludwig2015robust}.
This work also appears to be the first suggesting a hybrid model for the IVS combining a neural network component, and a standard model component.



\textbf{Paper structure.}
\Cref{sec:ivs} reviews background knowledge such as the implied volatility surface, the no-arbitrage conditions, and formulates the modeling problem.
We describe the methodology in~\Cref{sec:smoo}.
The numerical experiments and the empirical analysis can be found in~\Cref{sec:results}.
\Cref{sec:impact} discusses the broader impact. \Cref{sec:ccl} concludes.
More information on standard option pricing models, on implied volatility models, on the experiments, as well as additional results can be found in the online Appendix.


\section{Background and objectives} \label{sec:ivs}

\subsection{The implied volatility surface}

Let $\pi(K, \tau)$ denotes the market price of a call option with time to maturity $\tau>0$ and strike price $K\ge0$.
Without loss of generality, we assume that the initial stock price is $S$, and that the interest-rate $r$ and dividend yield $\delta$ are constants.
Denote by $C(\cdot)$ the Black-Scholes formula, namely
\begin{equation}\label{eq:BSMcp}
C(S, \, \sigma, \, r, \, \delta, \,K, \, \tau) = S^{-\delta \tau} \Phi(d_+) - \e^{-r\tau}K \Phi(d_-),
\end{equation}
where 
$
d_\pm = (\log(S/K) + (r-\delta)\tau)/(\sigma \sqrt{\tau}) \pm (1/2)\sigma\sqrt{\tau}.
$
More details on this formula can be found in Appendix~\ref{sec:app_bs}.
The main objective of this paper is modeling the implied volatility whose definition is given below.
\begin{definition}\label{def:ivs}
The \emph{implied volatility} $\iv(k,\tau)>0$ is given by the equation
\begin{equation} \label{eq:iv_def}
\pi(K,\tau) = C(S, \, \iv(k, \tau), \, r, \, \delta, \, K, \, \tau),
\end{equation}
with the (forward) log moneyness $k=\log(K/S\e^{(r-\delta)\tau})$.
The \emph{implied volatility surface} is given by $\iv(k,\tau)$ for $k\in\R$ and $\tau>0$.
\end{definition}

\textbf{Interpreting the IVS shape.}
For a fixed $\tau>0$,  $\iv(k,\tau)$ with $k\in\R$ defines a volatility smile.
If the smile has a \emph{U shape}, then the tail of the log return $\log(S_T/S_t)$ distribution are thicker than the tails of the Gaussian distribution, and vice versa.
If the smile exhibits a skew, then one side of the log return distribution is thicker than the other.
For example, if the left side of a smile, which is a slice of the surface for a fixed $\tau$, is steeper than the right side, then the log price is more likely to experience large losses than large gains.
The implied volatility surface provides a snapshot representation of valid option prices at a given time point.
Although option prices fluctuate significantly over time, the shape and level of the implied volatility surface is fairly stable and large movements indicate important changes in market conditions.


\subsection{Aribtrage-free surface} \label{sec:ao}

A static arbitrage is a static trading strategy that has
a value that is both zero initially and always greater than or equal to zero afterwards,
and a non-zero probability of having a strictly positive value in the future.
In other words, an arbitrage costs nothing to implement while only providing upside potential, that is,
it represents a risk-free investment after accounting for transaction costs.
Under the assumption that economic agents are rational, any such opportunity should be instantaneously exploited until the market is arbitrage free.
Therefore, option pricing models are designed in such a way that their call price surface $\pi(K,T)$ offers no possibility to implement such a strategy.
Standard static arbitrage opportunities are described in Appendix~\ref{sec:staticarb}.

The absence of arbitrage translates into \emph{constraints} on the call price surface $\pi(K,T)$, 
which in turn can be expressed as conditions that the implied volatility surface $\iv(k,\tau)$ must satisfy~\cite{roper2010arbitrage,gatheral2014arbitrage}.
To express those conditions, we define the \emph{total variance} of $\iv(k,\tau)$,
\begin{equation} \label{eq:siv_def}
  \omega(k,\tau)= \iv^2(k,\tau) \, \tau.
\end{equation}

\begin{proposition}{\citet[Theorem~2.9]{roper2010arbitrage}} \label{prop:noAO}
Let $S>0$, $r=\delta=0$, and $\omega:\R \times [0,\infty) \mapsto \R$.
Let $\omega$ satisfy the following conditions:
\begin{enumerate}[label={C\arabic*)}, noitemsep, topsep=0pt]
\item \label{cons:posi} (Positivity) for every $k\in\R$ and $\tau>0$,
$
\omega(k,\tau)>0.
$
\item \label{cons:vmat} (Value at maturity) for every $k\in\R$,
$
\omega(k,0)=0.
$
\item \label{cons:smoo} (Smoothness) for every $\tau>0$, $\omega(\cdot, \, \tau)$ is twice differentiable.
\item \label{cons:mono} (Monotonicity in $\tau$) for every $k\in\R$, $\omega(k, \, \cdot)$ is non-decreasing,
$
\ell_{\rm cal}(k,\tau) = \partial_\tau \omega(k,\tau) \ge 0,
$
where we have written $\partial_\tau$ for $\partial/\partial \tau$.
\item \label{cons:durr} (Durrleman's Condition) for every $\tau>0$ and $k\in\R$,
$$
\ell_{\rm but}(k,\tau) = \left( 1 - \frac{k \, \partial_k \omega(k,\tau)}{2 \omega(k,\tau)}\right)^2 - \frac{\partial_k \omega(k,\tau)}{4} \left(\frac{1}{\omega(k,\tau)} + \frac{1}{4}\right) + \frac{\partial^2_{kk}\omega(k,\tau)}{2} \ge 0,
$$
where we have written $\partial_k$ for $\partial/\partial k$ and $\partial_{kk}$ for $\partial^2/(\partial k\partial k)$
\item \label{cons:asym} (Large moneyness behaviour) for every $\tau>0$, $\sigma^2(k,\tau)$ is linear for $k\rightarrow \pm \infty$.
\end{enumerate} 
Then, the resulting call price surface is free of static arbitrage.
\end{proposition}

\labelcref{cons:posi} and~\labelcref{cons:vmat} are necessary conditions that any sensible model must satisfy.
As for~\labelcref{cons:smoo}, it is merely sufficient to prove an absence of arbitrage when~\labelcref{cons:mono},~\labelcref{cons:durr}, and~\labelcref{cons:asym} are also satisfied.
Note that, assuming~\labelcref{cons:smoo}, \labelcref{cons:mono} (respectively~\labelcref{cons:durr} and~\labelcref{cons:asym}) is satisfied if and only if the call price surface is free of calendar spread (butterfly) arbitrage~\cite{gatheral2014arbitrage}.
\labelcref{cons:asym} could be refined by imposing that $\sigma^2(k,\tau)/ \lvert k \rvert < 2$ when $k\rightarrow \pm \infty$ to guarantee the existence of higher order implied moments, see~\cite{lee2004moment,benaim2009regular}.

\begin{remark}
The Proposition~\ref{prop:noAO} is derived under the assumption that $r=\delta=0$ without loss of generality.
Indeed, the static no-arbitrage constraints have the same functional forms as in \ref{cons:mono}--\ref{cons:durr}--\ref{cons:asym} with non-zero parameters and the forward log moneyness $k$ defined in Definition~\ref{def:ivs}. 
\end{remark}

\subsection{Problem formulation}

\textbf{Data availability}.
In practice, market data may need to be validated.
This could be for a variety of reasons.
First, exchange traded securities have at least two quotes, a bid and a ask price, and there is no guarantee that the average prices are arbitrage-free.
Second, the observed prices may not be refreshed and thus not actionable, which may translate into notable input data noise.
In addition, market data is typically sparse away from the money and dense close to the money.
Indeed, far out of the money options are less likely to be exercised and are thus less likely to be used by their buyers.
There is typically more demand and supply for contracts that are around the money.

\textbf{Modeling objectives}.
The goal is to construct an IVS model $\sigma(k,\tau)$ that (I)
generates options prices that are in line with market data,
(II) is free of static arbitrage opportunities in the sense of~\Cref{prop:noAO},
and (III) generalizes to unobserved data regions in a controlled fashion.

 
\section{Methodology} \label{sec:smoo}

\subsection{Model and loss function} \label{sec:archlearn}

\textbf{Explanatory and target variables.}
At a given time we observe triplets $(\sigma_i, \, k_i, \, \tau_i)\in\datamkt$
where $\sigma_i$ is the market implied volatility (the target/response), and $(k_i, \tau_i)$ are the log moneyness and the time to maturity (the features/explanatory variables).
In addition, we complement the sample with synthetic pairs $(k_i, \, \tau_i)\in \datacalbut \cup\dataasy$ used to control the arbitrage opportunities and the model asymptotic behavior (see~\Cref{sec:noarb}).

\textbf{Implied volatility model.}
Our model for the total variance and implied volatility is given by
\begin{align}\label{eq:iv_ann}
\wth(k,\tau) &= \wnn\big(k,\tau;\theta_1\big) \times \wpr\big(k,\tau;\theta_2\big)  \quad \mbox{and} \quad
\ivth(k,\tau) = \sqrt{\wth(k,\tau)/\tau}
\end{align} 
for the parameters $\theta=\{\theta_1, \theta_2\}$, where
$\wnn$ and $\wpr$ are the NN and prior models described below.

\textbf{Prior model.}
$\wpr:\R^2\mapsto\R$ is a prior model with parameters $\theta_2$.
An implied volatility model without prior is obtained by setting $\wpr\equiv 1$ so that $\wth \equiv \wnn $. 
The prior model choice is useful to ensure that the model generalization is compliant with a prescribed preferred behavior.
Essentially, the BS prior is a parameter-free model matching the implied volatility's at-the-money (ATM) term-structure.
As for the SSVI prior, it improves on the BS model by capturing both the smile and the skew of the surface.
Both models are described in Appendix~\ref{sec:bsprior} and~\ref{sec:ssviprior} respectively.

\textbf{NN model.}
$\wnn:\R^2\mapsto\R$ is a standard feedforward multilayer neural network, namely
\begin{align}
\wnn(k, \tau;\theta_1) &=  \comp_{i=1}^{n+1} f_i^{W_i, b_i} (k, \tau) \mbox{ with } 
f_i(x) = \begin{cases} g_i(W_i \, x + b_i) & i< n + 1
\\ \alpha \left(1 + \tanh \left(W_{n+1} \, x + b_{n+1} \right)\right) & i=n+1 \end{cases} \label{eq:ffnn_def2}
\end{align}
with $g_i$ an activation function, $\theta_1=\{W_1,\,b_1,\,W_2,\,b_2,\dots, \alpha\}$ the set of weight matrices and bias vectors, $\alpha$ a scaling parameter letting $\wnn$ take values in $[0, \alpha]$, and $n$ the number of hidden layers.
The functional choice $1+\tanh$ in the last layer is not critical, yet it appeared to perform slightly better than the sigmoid function on a limited number of trials.
Note that one can initialize the NN to take output value one, that is no NN correction, by setting $\alpha=1$ and $W_{n+1}=b_{b+1}=0$.

\textbf{Loss function.}
We fit the network parameters and prior parameters $\theta$ by minimizing the loss function
\begin{equation}\label{eq:loss}
\Lcal (\theta) = \Lcal_0(\theta) + \sum_{j=1}^6 \; \lambda_j \, \Lcal_{{\rm C} j}(\theta)
\end{equation}
where the term $\Lcal_0(\theta)$ is a prediction error cost, the terms $\Lcal_{{\rm C} j}(\theta)$ for $j=1,\dots,6$ materialize soft constraints aiming to ensure that the shape of $\{\wth(k, \tau);\, (k, \tau) \in\R\times\R_+\}$ is indeed a sensible implied volatility surface, and $\lambda_i$ for $j=1,\dots,6$ are the corresponding penalty weights.
Note that some parameters of the prior model may also be calibrated.

We let the prediction error be the sum of the root-mean-squared-error (RMSE) and the mean-absolute-percentage-error (MAPE),
\[
\Lcal_0(\theta) = \sqrt{1/|\datamkt|\sum_{(\sigma_i,k_i, \tau_i)\in\datamkt} (\sigma_i - \ivth(k_i, \tau_i))^2} + \;1/|\datamkt| \sum_{(\sigma_i,k_i, \tau_i)\in\datamkt}  \vert \sigma_i - \ivth(k_i, \tau_i) \vert/\sigma_i,
\]
so as to penalize both absolute and relative errors.
The IV values far out of the money can take values significantly larger than at the money.
We found that the above loss function results in general a balanced allocation of prediction errors across the different moneynesses.

\begin{remark}[Soft versus hard constraints]
An alternative approach to impose shape constraints on the mapping $\wth$ is to hard-wire them into the neural network architecture, as in~\cite{dugas2001incorporating} for example.
However, hard constraints are difficult to impose on multilayer neural networks, may reduce the neural network's flexibility, and may lead to more challenging leaning routines, see~\cite{marquez2017imposing}.
\end{remark}

\subsection{No-arbitrage conditions and synthetic grid} \label{sec:noarb}

We explain how each of the constraints/conditions in~\Cref{prop:noAO} can be handled either by refining the architecture of the neural network, or by adding a penalty term to the loss function~\eqref{eq:loss}.
Note that the conditions \textbf{\ref{cons:posi}--\ref{cons:vmat}} are in principle satisfied by design of the NN.
The mapping $\wth$ is twice differentiable as long as the activation functions $g_i$ and $g_{n+1}$ (as well as the prior model) are twice differentiable, in which case\textbf{ \ref{cons:smoo}} is satisfied.
An example of valid activation function is the SoftPlus given by $\ln(1+\exp(x))$\footnote{We conjecture that one could also use ReLU activation functions with adjusted constraint conditions.}.
Hence we set $\Lcal_{\rm C1} \equiv \Lcal_{\rm C2}\equiv\Lcal_{\rm C3}\equiv 0$.

As for the other three constraints, we control for
\begin{itemize}
\item calendar arbitrages with $\Lcal_{\rm C4} (\theta) = 1/|\datacalbut| \sum_{(k_i,\tau_i)\in\datacalbut} \max\left(0, \, - \ell_{\rm cal}(k_i,\,\tau_i) \right),$
\item butterfly arbitrages with $\Lcal_{\rm C5} (\theta) = 1/|\datacalbut| \sum_{(k_i,\tau_i)\in\datacalbut}  \max\left(0, \, -\ell_{\rm but}(k_i,\,\tau_i)\right)$,
\item and the asymptotic behavior with $\Lcal_{\rm C6} (\theta) = 1/|\dataasy| \sum_{(k_i,\tau_i)\in\dataasy} \left\lvert \partial^2 \wth(k_i,\tau_i)/\partial k \partial k \right\rvert$,
\end{itemize}
where 
\begin{align*}
\datacalbut &= \big\{(k,\tau) \, : \, k\in \big\{ x^3 \,:\, x \in \big\lbrack -(-2 k_{\min})^{1/3}, \, (2k_{\max})^{1/3} \big\rbrack_{100} \big\}, \,\tau \in \Tcal\big\} \\
\dataasy &= \big\{(k,\tau) \, : \, k\in \big\{6k_{\min}, \, 4k_{\min} , \, 4k_{\max}, \, 6k_{\max} \big\}, \,\tau \in \Tcal\big\}\\
\Tcal &= \big\{ \exp(x) \,:\, x \in \big\lbrack \log(1/365), \, \max( \log(\tau_{\max}+1)) \big\rbrack_{100} \big\}
\end{align*}
 $k_{\min} = \min(\datamkt^k)$, $k_{\max} = \max(\datamkt^k)$, $\tau_{\max} = \max(\datamkt^\tau)$, $\lbrack a,b \rbrack_{x}$ indicates an equidistant set of $x$ points between $a$ and $b$, and $\datamkt^\tau$ and $\datamkt^k$ are the sets of unique time to maturity and forward log moneyness in $\datamkt$.
Note that it should always be that $\min(\datamkt^k)<0$ and $\max(\datamkt^k)>0$.
The motivation for the above transformations is to obtain a denser grid around the money and for short maturities.
The particular parametric choices for the above sets do not appear critical as long as they are sufficiently dense and cover the regions of interest.


Note that the loss $\Lcal_{\rm C6}$ guarantees that $\omega$ is asymptotically linear and should in principle be refined to control the coefficient value to be less than 2, as discussed in~\cite{lee2004moment}.
However $\Lcal_{\rm C6}$ has better training performance and proved to be compliant with~\ref{cons:asym} in our post training verifications.
Furthermore, we only work with finite moneyness in practice and imposing an asymptotic condition on large but finite moneyness values may unnecessarily constraint the model.

\subsection{Model training}

The training procedure and the parameters choice are described in details in Appendix~\ref{sec:train}.


\section{Results} \label{sec:results}

The first part presents results on synthetic data where it is easier to study and compare model predictions.
The second part presents results on real-world data to modeling implied volatility surfaces extracted from S\&P500 options prices.
The synthetic and market data are described in Appendix~\ref{sec:datasynt} and~\ref{sec:datamkt} respectively.
We always set $\lambda_4=\lambda_5=\lambda_6=\lambda$ for some $\lambda\ge0$.

\subsection{Numerical experiments} \label{sec:numexp}

\begin{figure}
\centering
\includegraphics[width=\textwidth]{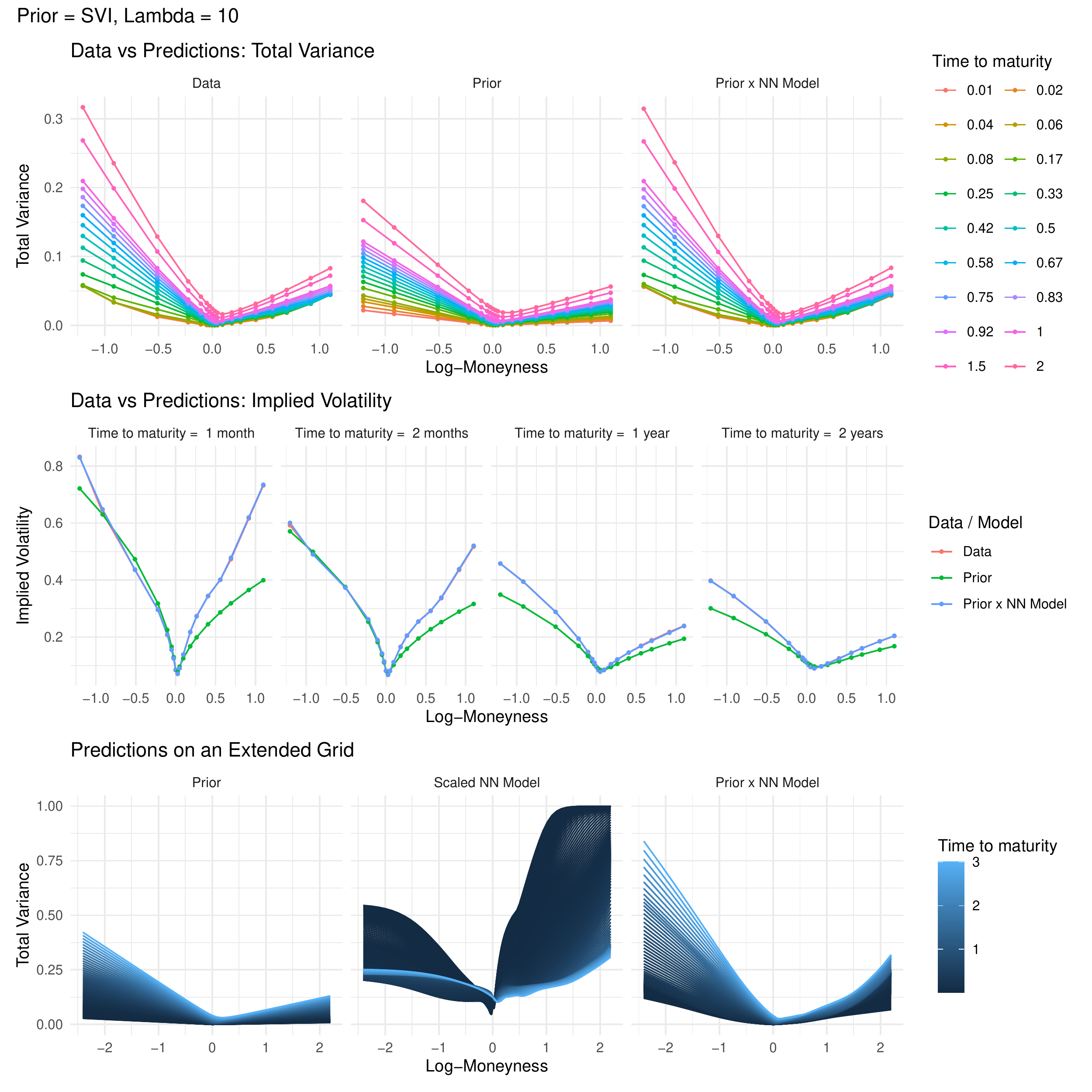}
\caption{\label{fig:smoothsynt} Synthetic data and trained model predictions for a specific configuration (scenario 12).}
\end{figure}

\textbf{Smoothing behavior.}
\Cref{fig:smoothsynt} displays the trained model on synthetic data with a strong penalty $\lambda=10$.
A dot indicates the positioning $(k,\tau)$ of an observation used for training on figures in the top row.
The figures on the first and second rows show that $\ivpr$ does not succeed to reproduce the training data, while $\ivth$ does so perfectly.
The figures in the last row display $\ivpr$ and $\ivth$ on an extended grid of log moneyness, and also shows the NN correction before the scaling by $\alpha$ (center figure).
As $\ivpr$ fits at the money (ATM) term structure, $\ivth$ makes little correction to it on the region $k\approx0$.
On the other hand, $\ivth$ makes important corrections to $\ivpr$ for out of the money options.

The same figure for additional scenarios ($\lambda$ value, prior model, and data limits) can be found in Appendix~\ref{sec:smoothsynt}.
These additional results show that the model generalization is significantly better with a SSVI prior and with soft-constraints.
Indeed, the IVS tends to flatten for large log moneyness values with a BS prior, whereas the present data is V shaped.
Also, it is shown that the absence of soft-constraints can translate into non-realistic and possibly absurd out of sample predictions.



\textbf{Losses and convergence.}

\begin{figure}
\centering
\includegraphics[width=\textwidth]{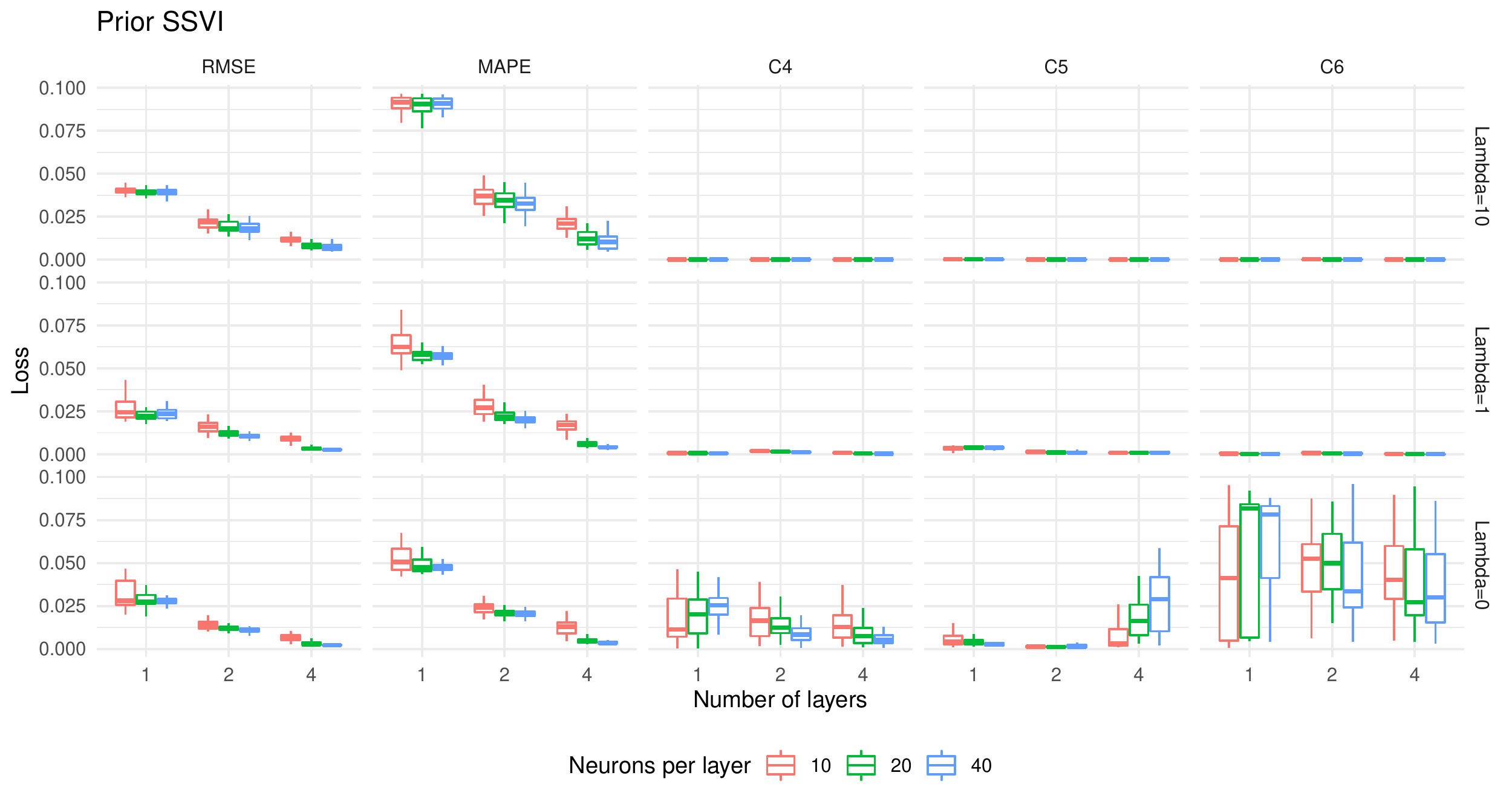}
\caption{\label{fig:convsvi} Losses for different number of layers, neurons per layer, and penalty value.}
\end{figure}

\Cref{fig:convsvi} displays statistics on the losses in~\eqref{eq:loss} for trained models with different number of layers, neurons per layer, and penalty value $\lambda$.
A total of 50 models with different random seeds for parameter initialization have been trained for 5'000 epochs for each configuration.
Recall that, with $\lambda=x$, the weight on the soft constraints is $x$ times larger than the loss on the predictions errors.
We see that increasing the number of layers or of neurons per layers generally decreases the RMSE and MAPE.
We also see that the arbitrage opportunities vanish as $\lambda$ increases.
The models prediction errors typically worsen as $\lambda$ increases, but the effect appears stronger for the lesser flexible models.
We also observe that, when the number of layers is smaller, increasing the neurons per layer without enforcing more strictly the constraints can generate more severe arbitrage opportunities.
But such opportunities disappear completely when the absence of arbitrage penalty is given a higher weight.
Additional results can be found in Appendix~\ref{sec:convergence} for the BS prior where,
in particular, we observe that the prediction errors are worse for this choice of prior model.

\textbf{Additional results.}
In Appendix~\ref{sec:dugasetal} we show that the NN model suggested in~\cite{dugas2001incorporating} cannot capture the IVS on its full range of maturities and moneynesses for the synthetic data.
We also illustrate the impact of increasing arbitrage opportunities on the losses in Appendix~\ref{sec:numpexpao}.

\subsection{Empirical results} \label{sec:empres}

\begin{figure}
\centering
\includegraphics[width=\textwidth]{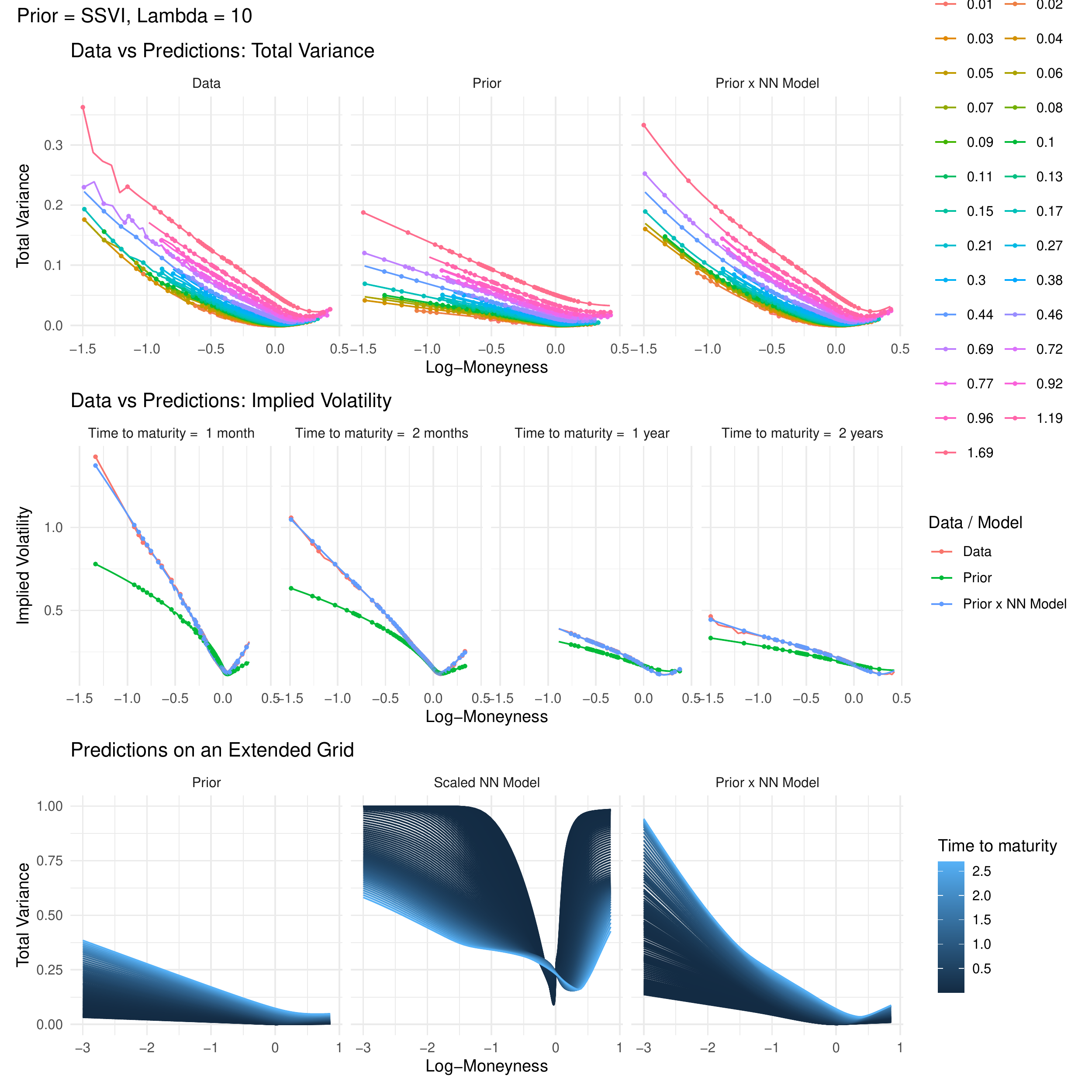}
\caption{\label{fig:smoothmkt} Market data and trained model predictions for a specific configuration (scenario 12).}
\end{figure}

\textbf{Smoothing market data.}
\Cref{fig:smoothmkt} displays S\&P500 options data and trained model predictions with a strong penalty $\lambda=10$ for the April 13-th 2018.
This market data contains several thousands observations, and the corresponding IVS has a fairly complex shape compared to the previous synthetic data. 
The first two rows highlight that $\ivpr$ fails to reproduce the market data, but that $\ivth$ is able to make accurate predictions.
Notice that, because of data noise, the target IVS seem not to be arbitrage-free as some total variance slices cross each others, see upper-left figure.
Yet, the predictions generated by $\ivth$ are perfectly smooth and the slices do not cross.
The same figure for additional scenarios can be found in Appendix~\ref{sec:smoothmkt}.

\textbf{Backtests.}
We perform the following exercise for each day in the sample.
First, we split the daily sample into a training and a testing set.
Second, we fit the model on the training set and evaluate its performance on the testing set.
We use two different configurations for training and testing.
In the interpolation setting, for each maturity, we randomly select half of the contracts.
As such, we also sample options that are far out or in the money for training, and the testing error represents the approximation error for the range of moneyness that are actually observed.
In the extrapolation setting, for each maturity, we select half of the contracts whose log moneyness is between the 10\% and 90\% of the log moneyness in the corresponding slice.
This second filter therefore contains more observations around the money.
Thus, we do not select options that are far out or in the money for training, and the testing error measures how well our model extrapolates.
Finally, we again use three values for $\lambda$ in order to study how the arbitrage-related penalties affect the results.

In~\Cref{tab:bktSSVI} we present our results for model trained daily on the S\&P500 options data between January and April 2018, and for the subprime crisis period between September and December 2008, with a SSVI prior.
First, we describe the RMSE and MAPE.
As expected, training errors are generally below testing errors.
Increasing the no-arbitrage penalty $\lambda$ leads to worse RMSE and MAPE metrics on the training set.
However, larger $\lambda$ also implies similar or even less RMSE and MAPE metrics on the testing set.
Note that the RMSE for testing set of the extrapolation appears to be large compared to the MAPE.
The extrapolation errors are most of the time larger than interpolation errors.
This suggest that the high RMSE value is likely to be caused by deep out of the money options with large IV values.
Regarding \ref{cons:mono}--\ref{cons:asym}, we see that the models resulting from $\lambda=1,10$ are essentially arbitrage-free.
As for the model resulting from no enforcement of the constraints (i.e., $\lambda=0$), it generates arbitrage opportunities both in interpolation and extrapolation.
Note that 99+ indicates a value larger than 99 which can happen for arbitrage losses when $\lambda=0$.

Additional results for the BS prior, as well as for baselines models (Bates and SSVI models), can be found in Appendix~\ref{sec:backtests}.
Our approach surpasses the baseline models in terms of prediction accuracy, which should not be a surprise.
Indeed, in Section~\ref{sec:numexp} we showed that $\ivth$ could reproduce the IVS from a typical Bates model, hence it is likely to be at least as flexible.
In addition, $\ivth$ with a SSVI prior $\ivpr$ is likely to perform better than the underlying baseline SSVI model.

\begin{table}
\vspace{-0.5cm}
\caption{\label{tab:bktSSVI}Backtesting results for the SSVI prior (quantiles in \%, Jan-Apr 2018 / Sep-Dec 2008)}
\centering
\resizebox{\textwidth}{!}{
\begin{tabular}[t]{cccccccccccccc}
\toprule
\multicolumn{1}{c}{ } & \multicolumn{1}{c}{ } & \multicolumn{6}{c}{Interpolation} & \multicolumn{6}{c}{Extrapolation} \\
\cmidrule(l{3pt}r{3pt}){3-8} \cmidrule(l{3pt}r{3pt}){9-14}
\multicolumn{1}{c}{ } & \multicolumn{1}{c}{ } & \multicolumn{3}{c}{Train} & \multicolumn{3}{c}{Test} & \multicolumn{3}{c}{Train} & \multicolumn{3}{c}{Test} \\
\cmidrule(l{3pt}r{3pt}){3-5} \cmidrule(l{3pt}r{3pt}){6-8} \cmidrule(l{3pt}r{3pt}){9-11} \cmidrule(l{3pt}r{3pt}){12-14}
Loss & $\lambda$ & $q_{05}$ & $q_{50}$ & $q_{95}$ & $q_{05}$ & $q_{50}$ & $q_{95}$ & $q_{05}$ & $q_{50}$ & $q_{95}$ & $q_{05}$ & $q_{50}$ & $q_{95}$\\
\midrule
 & 10 & 0.4 / 0.7 & 0.5 / 3.1 & 1.3 / 19.9 & 0.4 / 0.9 & 0.5 / 3.3 & 1.4 / 18.8 & 0.2 / 0.3 & 0.3 / 1.1 & 0.4 / 3.5 & 2.2 / 2.7 & 5.0 / 6.3 & 8.0 / 12.0\\
\cmidrule{2-14}
 & 1 & 0.3 / 3.0 & 0.4 / 6.8 & 1.0 / 13.9 & 0.4 / 2.9 & 0.5 / 7.1 & 1.3 / 13.5 & 0.2 / 1.9 & 0.2 / 4.4 & 0.4 / 10.6 & 3.7 / 5.0 & 6.6 / 10.6 & 11.7 / 20.5\\
\cmidrule{2-14}
\multirow{-3}{*}{\centering\arraybackslash RMSE} & 0 & 0.2 / 0.4 & 0.3 / 1.3 & 0.5 / 4.0 & 0.3 / 1.6 & 0.5 / 3.5 & 4.8 / 10.8 & 0.2 / 0.3 & 0.2 / 0.9 & 0.3 / 3.3 & 3.1 / 5.5 & 7.5 / 11.6 & 18.1 / 26.7\\
\cmidrule{1-14}
 & 10 & 0.5 / 0.9 & 0.7 / 2.1 & 1.2 / 17.4 & 0.5 / 1.1 & 0.8 / 2.4 & 1.2 / 18.5 & 0.4 / 0.5 & 0.6 / 0.9 & 0.9 / 2.1 & 1.2 / 2.4 & 1.7 / 3.3 & 2.4 / 6.5\\
\cmidrule{2-14}
 & 1 & 0.5 / 4.0 & 0.6 / 8.2 & 1.2 / 12.2 & 0.5 / 4.3 & 0.7 / 8.1 & 1.3 / 12.9 & 0.3 / 2.2 & 0.5 / 6.3 & 0.9 / 11.0 & 1.5 / 5.6 & 2.3 / 9.7 & 3.3 / 13.1\\
\cmidrule{2-14}
\multirow{-3}{*}{\centering\arraybackslash MAPE} & 0 & 0.4 / 0.5 & 0.6 / 1.0 & 0.9 / 1.8 & 0.5 / 1.2 & 0.7 / 1.9 & 0.9 / 3.2 & 0.3 / 0.3 & 0.5 / 0.7 & 0.8 / 1.6 & 1.5 / 4.3 & 2.2 / 7.4 & 4.8 / 14.3\\
\cmidrule{1-14}
 & 10 & 0.0 / 0.0 & 0.0 / 0.0 & 0.0 / 0.2 & 0.0 / 0.0 & 0.0 / 0.0 & 0.0 / 0.5 & 0.0 / 0.0 & 0.0 / 0.0 & 0.0 / 0.0 & 0.0 / 0.0 & 0.0 / 0.0 & 0.0 / 14.1\\
\cmidrule{2-14}
 & 1 & 0.0 / 0.0 & 0.0 / 0.0 & 0.2 / 0.1 & 0.0 / 0.0 & 0.0 / 0.0 & 0.2 / 0.1 & 0.0 / 0.0 & 0.0 / 0.0 & 0.0 / 0.1 & 0.0 / 0.0 & 0.0 / 0.0 & 0.1 / 0.1\\
\cmidrule{2-14}
\multirow{-3}{*}{\centering\arraybackslash C4} & 0 & 1.6 / 1.5 & 18.0 / 9.8 & 99+ / 99+ & 1.6 / 1.2 & 40.7 / 12.1 & 99+ / 99+ & 0.3 / 1.1 & 2.4 / 3.7 & 44.2 / 83.7 & 0.4 / 1.0 & 2.8 / 4.3 & 46.7 / 30.4\\
\cmidrule{1-14}
 & 10 & 0.0 / 0.0 & 0.0 / 0.0 & 0.0 / 0.1 & 0.0 / 0.0 & 0.0 / 0.0 & 0.0 / 0.1 & 0.0 / 0.0 & 0.0 / 0.0 & 0.0 / 0.0 & 0.0 / 0.0 & 0.0 / 0.0 & 0.0 / 1.2\\
\cmidrule{2-14}
 & 1 & 0.0 / 0.0 & 0.0 / 0.0 & 0.1 / 0.0 & 0.0 / 0.0 & 0.0 / 0.0 & 0.5 / 0.0 & 0.0 / 0.0 & 0.0 / 0.0 & 0.0 / 0.0 & 0.0 / 0.0 & 0.0 / 0.0 & 0.5 / 0.0\\
\cmidrule{2-14}
\multirow{-3}{*}{\centering\arraybackslash C5} & 0 & 2.6 / 99+ & 72.1 / 99+ & 99+ / 99+ & 2.6 / 99+ & 63.8 / 99+ & 99+ / 99+ & 2.8 / 99+ & 24.2 / 99+ & 99+ / 99+ & 1.7 / 99+ & 19.0 / 99+ & 99+ / 99+\\
\cmidrule{1-14}
 & 10 & 0.0 / 0.0 & 0.0 / 0.0 & 0.0 / 0.2 & 0.0 / 0.0 & 0.0 / 0.0 & 0.0 / 0.6 & 0.0 / 0.0 & 0.0 / 0.0 & 0.0 / 0.0 & 0.0 / 0.0 & 0.0 / 0.0 & 0.0 / 44.3\\
\cmidrule{2-14}
 & 1 & 0.0 / 0.0 & 0.0 / 0.0 & 0.0 / 0.0 & 0.0 / 0.0 & 0.0 / 0.0 & 0.6 / 0.0 & 0.0 / 0.0 & 0.0 / 0.0 & 0.0 / 0.1 & 0.0 / 0.0 & 0.0 / 0.0 & 0.0 / 0.0\\
\cmidrule{2-14}
\multirow{-3}{*}{\centering\arraybackslash C6} & 0 & 0.8 / 0.0 & 48.2 / 1.7 & 99+ / 99+ & 0.8 / 0.0 & 78.4 / 0.4 & 99+ / 99+ & 0.3 / 0.1 & 12.0 / 7.5 & 99+ / 99+ & 0.0 / 0.0 & 2.4 / 0.0 & 99+ / 99+\\
\bottomrule
\end{tabular}
}
\end{table}

\textbf{Additional results.}
In Appendix~\ref{sec:pdflocvol} we display the risk-neutral density and the local volatility corresponding to the scenario 12.
The experience on Apple options in Appendix~\ref{sec:wshape} shows that our approach can fit complex shape surfaces, such as W-shaped smiles.
We trained a model on the absolute IV errors weighted by the IV spread in Appendix~\ref{sec:spreadw} which appears to converge faster but fails to learn the IVS shape far out of the money.
The computational times for the empirical experiments are reported in Appendix~\ref{sec:time} and appear satisfying despite the experimental nature of the implementation.


\section{Conclusion}\label{sec:ccl}

We described a flexible methodology the price financial derivatives in an economically sensible way.
This is achieved by modeling the implied volatility surface with a multilayer neural network and shaping it by penalizing the loss.
We validate our approach with various numerical and empirical applications.
The presented approach could be used as a building block to construct arbitrage-free models for multiple stocks, and for the IVS dynamics, as further discussed in Appendix~\ref{sec:discussion}.
Additionally, the loss function could be improved by avoiding to penalize models for predictions inside the spread, or using vega-weighting, as is commonly done by practitioners.

\section*{Broader Impact}\label{sec:impact}

Financial markets play a central role in our economy, and our work could be used to generalize in a robust way the information available to participants.
Options are actively traded securities that can be used for multiple reasons such as protecting pension portfolios against future losses, hedging against future fluctuations in crop prices for agricultural farmers, or speculation for hedge funds.
Options are also used by economist and financial expert to extract market implied sentiment measures such as the VIX ``fear'' index.
The financial risk created by options is often held by financial intermediaries, such as banks or brokers, which need appropriate tools to monitor and control their financial risk.

\begin{ack}
The authors would like to thank Michael Roper for providing detailed comments and suggestions, as well as Serge Kassibrakis, Charles-Albert Lehalle, Johannes Wiesel, and participants at the 2019 SIAM conference on Financial Mathematics and Engineering in Toronto for helpful discussions.

The statements and opinions expressed in this article are those of the authors and do not represent the views of UBS (and/or any branches) and/or their affiliates.
\end{ack}

\bibliographystyle{rusnat}
\bibliography{bibliography}

\newpage
\begin{appendices}

\section{Background on option pricing models} \label{sec:bs}

\subsection{Literature review}\label{sec:litrev}


The assumption of constant volatility in the Black-Scholes-Merton model has long been challenged empirically and various stochastic volatility models have been developed to tackle its limitations.
Some examples are the Hull-White~\cite{hull1987pricing}, the Stein-Stein~\cite{stein1991stock}, the Heston~\cite{heston1993closed}, the Variance-Gamma~\cite{madan1998variance}, the normal inverse Gaussian~\cite{barndorff1997processes}, the CGMY~\cite{carr2002fine}, the 4/2~\cite{grasselli2017}, the Jacobi~\cite{ackerer2018jacobi}, rough Heston~\cite{el2019characteristic}, and affine Volterra~\cite{jaber2017affine} models.

Albeit the development of more flexible models for stock prices, their statistical flexibility remained limited and they may be computationally too costly to calibrate for some real-world applications.
For these reasons, parametric and nonparametric approaches have been developed aiming to interpolate, and sometimes to extrapolate, the implied volatility surface.
These approaches includes the stochastic volatility inspired (SVI) and surface extension (SSVI) of~\cite{gatheral2004parsimonious, gatheral2014arbitrage}, and the smoothing spline techniques of~\cite{fengler2009arbitrage, corlay2016b}, 
among many others.

Several shallow neural networks approaches have also been developed to smooth option prices directly.
\cite{dugas2001incorporating} constructed a one hidden layer neural network monotonic or convex in its input coordinate, and taking only positive values.
However, the construction is specific rendering it impossible to extend to multilayer neural networks, it performs poorly with both short and long maturities, and do not prevent all forms of static arbitrage opportunities.
Recently, this model has been extended in a PhD thesis~\cite{zheng2018machine} by adding a gated unit layer linking the input to multiple models \`a la Dugas.
\cite{ludwig2015robust} proposes a one hidden layer approach to model the implied total variance and his approach includes multiple ad-hoc rules.
For examples, 
the extrapolation behavior to unobserved areas of the implied volatility surface is controlled for by adding discretionary data points, 
the training procedure is restarted until 25 neural nets are found to be arbitrage-free at selected strikes and maturities,
the final implied volatility surface is obtained by aggregating over the best three models,
and so on.
The sigmoid-based approach of~\cite{itkin2015sigmoid} to model the implied volatility smile is closely related to a neural network approach.

On a broader note, the financial applications of neural networks are booming as a consequence of the progress made in deep learning and of the availability of specialized software and hardware.
They have for examples been used in~\cite{liu2019neural,liu2019pricing} to speed-up the pricing and calibration of options in stochastic volatility models, and in~\cite{buehler2019deep} to approximate optimal but intractable option hedging strategies with market frictions.

\subsection{The Black-Scholes (BS) formula} \label{sec:app_bs}

In the BS model, the dynamics of the stock price $S_t$ under the risk-neutral measure is given by
\begin{equation}\label{eq:BSM}
dS_t = (r-\delta)S_t dt + \sigma S_t dW_t
\end{equation}
for some constants $r\in\R$, $\delta\ge0$, and $\sigma>0$, and where $W_t$ is a standard Brownian motion.
Let $V_t$ denotes the price of a derivative at time $t$, then it satisfies the following partial differential equation,
\begin{equation}\label{eq:BSM_PDE}
0 = \frac{\partial V}{\partial t} + \frac{1}{2} \sigma^2 S^2 \frac{\partial^2 V}{\partial S^2} + (r-\delta)S \frac{\partial V}{\partial S} - rV.
\end{equation}
Consider the call option payoff $(x-K)^+$ with strike $K$ and maturity $T$.
Solving the PDE~\eqref{eq:BSM_PDE} with boundary condition $V_T=(S_T-K)^+$ with $S_t=S$ gives the following formula for the time-$t$ call option price $C$,
\begin{equation}
C(S, \, \sigma, \, r, \, \delta, \,K, \, \tau) = S^{-\delta \tau} \Phi(d_+) - \e^{-r\tau}K \Phi(d_-),
\end{equation}
where $\tau = T-t$,
$
d_\pm = (\log(S/K) + (r-\delta)\tau)/(\sigma \sqrt{\tau}) \pm (1/2)\sigma\sqrt{\tau}
$
and $\Phi$ is the standard Gaussian CDF.

The dynamics of stock prices in the real world do not follow a geometric Brownian motion. 
Empirically validated stylized fact of stock log returns are, for examples, stochastic volatility and leverage effect which are not capture by~\eqref{eq:BSM}.
Despite its shortcomings, the BS model remains extremely popular in practice for its simple pricing formula, and the modeling complexity is moved to the input volatility parameter $\sigma$.
Hence, if one understands the model and its limitations, the BS formula can be used as a Rosetta Stone to analyze market prices.

\subsection{Static arbitrage opportunities} \label{sec:staticarb}

\begin{figure}
\centering
\includegraphics[width=0.5\textwidth]{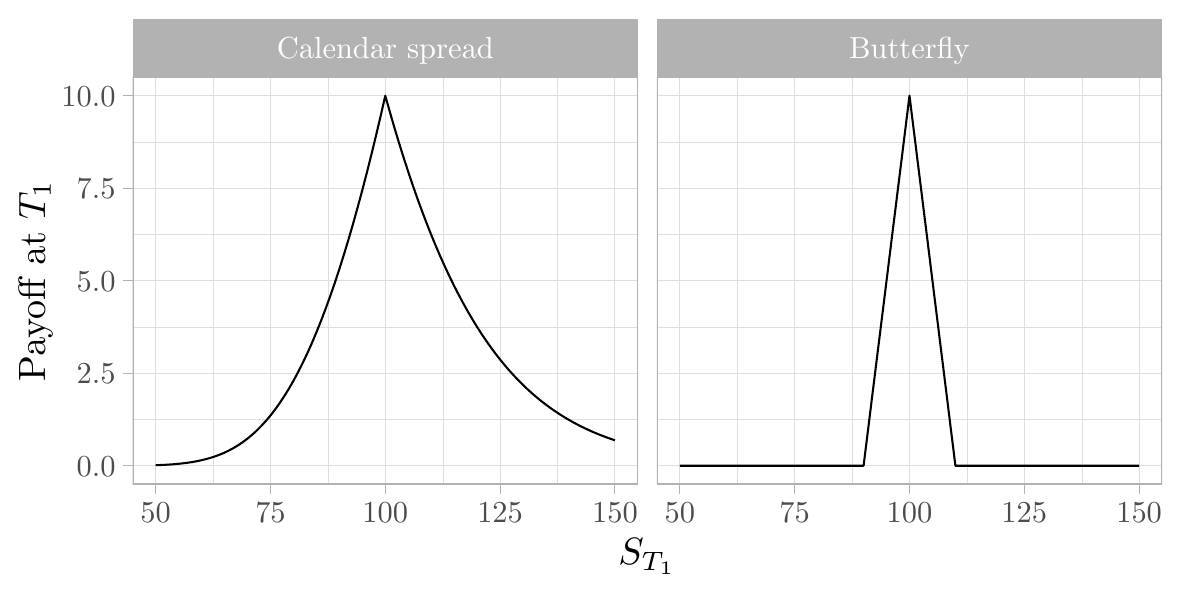}%
\caption{Payoffs of the calendar spread and butterfly as a function of the underlying asset price.
  For the calendar spread, $T_2 - T_1 = 1$, $K = 100$, and $\sigma = 0.25$. For the butterfly, $K_1 = 90$, $K_2 = 110$, and $K_3 = 100$.   \label{fig:payoffs}}
\end{figure}

One can show that $\pi(K,T)$ is arbitrage-free if and only if it is free of calendar spread arbitrage and each time slice if free of butterfly arbitrage.
A \emph{calendar spread} is a strategy where one buys a call with a given maturity $T_1$ and sells another call with maturity $T_2$, both using the same strike, and where $T_1 > T_2$.
At $T_1$, the value of the short call is $-\max(S_{T_1} - K, 0)$, whereas that of the long call, $\pi(K, T_2 - T_1)$, is always greater.
A \emph{butterfly} is a strategy where one buys two calls with strikes $K_1 < K_2$, and sells two other calls with strike $K_3 = (K_1 + K_2)/2$, but the same maturity.
In~\Cref{fig:payoffs}, we show the payoffs for each of the two strategies at $T_1$.
Since the payoffs are always positives, they must have a nonzero initial price, for the market would otherwise allow for arbitrage opportunities.

\subsection{Financial engineering} \label{sec:fineng}

The implied volatility surface plays a central role in the financial engineering toolbox.
Thanks to automatic differentiation, the model in~\eqref{eq:iv_ann} can also be used to derive the local volatility surface as well as the risk neutral density, for examples.
We refer to~\cite[Chapter~1]{gatheral2011volatility} for more background.

\textbf{Risk neutral density.}
Options are sometimes used to extract forward looking market sentiment indicators.
For examples, the VIX and the SKEW indices\footnote{See \url{http://www.cboe.com/VIX} and \url{http://www.cboe.com/SKEW}.} derive from the S\&P500's mean, variance, and skewness as implied by option prices.
In the present framework, we can reconstruct the entire fitted stock price density $p(x,\tau)$ at any future time $\tau$,
\begin{equation*}
p(x,\tau) = \frac{\partial^2 C }{ \partial K^2} \Big\lvert_{K=x} = \e^{r\tau} \Phi(d_+) \frac{\ell_{\rm but}(k,\tau)}{\sqrt{\omega (k, \tau)}}\Big\lvert_{K=x}.
\end{equation*}

\textbf{Local volatility.} 
The pricing of exotic and path-dependent options is often carried out using stochastic models equipped with a deterministic functional component that must be calibrated.
This is the so-called local stochastic volatility function $\lv(k,\tau)$ with log moneyness $k$ which is given by, 
\begin{equation*}
\lv(k,\tau)^2 
 = \frac{\frac{\partial C}{\partial T} + (r - \delta)K \frac{\partial C}{\partial K} + \delta C}{ \frac{1}{2} K^2 \frac{\partial^2C}{\partial K^2}} = \frac{ \frac{\partial w}{\partial T} }{1 - \frac{k}{w} \frac{\partial w}{\partial k}+  \frac{1}{4}\left( -  \frac{1}{4} +  \frac{1}{w} +  \frac{k^2}{w^2} \right) \left(\frac{\partial w}{\partial k}\right)^2 + \frac{1}{2}\frac{\partial^2 w}{\partial k^2} }.
\end{equation*}

\section{Prior and baseline models}\label{sec:models}

\subsection{Black-Scholes (BS) model} \label{sec:bsprior}

The BS model is the simplest possible prior model.
In its most standard form, one would define $\wpr^{\rm bs}(k,\tau) = \sigma^2 \tau$ for some volatility $\sigma$.
But, empirically, the at-the-money (ATM) total variance is not learn in $\tau$, and its term-structure can be inferred directly from market prices (see Appendix~\ref{sec:addit}).
As a result, we redefine the BS prior as
\begin{align*}
\wpr^{\rm bs}(k,\tau) & = \watm(\tau)
\end{align*}
where $\watm$ is described in Appendix~\ref{sec:addit}.
Hence, this prior can be inferred directly from market prices by interpolating/extrapolating the ATM total variance.

\subsection{Surface Stochastic Volatility Inspired (SSVI)} \label{sec:ssviprior}

The SVI parametrization for a volatility slice (single maturity) has been extended to the entire surface (SSVI) in~\cite{gatheral2014arbitrage}.
We implemented the following version with a power-law parameterization of the function $\phi$
\begin{align*}
\wpr^{\rm ssvi}(k,\tau) & = 
\frac{\watm(\tau)}{2} \big(1 + \rho\, \phi(\watm(\tau)) k + \sqrt{(\phi(\watm(\tau))k + \rho)^2 + 1 - \rho^2}\Big) \\
\phi(x) & =
\frac{\eta}{x^\gamma (1 + x)^{1 - \gamma}}
\end{align*}
for some parameters $\rho\in(-1,1)$, $\lambda\in(0,1)$, $\eta>0$, and where $\watm$ is the at the money term-structure of the IVS as described in Appendix~\ref{sec:addit}.
We also implemented the Heston-like parametrization for $\phi$ but the results were worse and are therefore not included.
A generalization of the SSVI parametrization is given in~\cite{guo2016generalized}.

\subsection{Stochastic volatility (SV) model} \label{sec:SVJ}

The stock price dynamics in the Bates model~\cite{bates1996jumps} is given by
\begin{align*}
dS_t/S_t &= (r - \delta)dt+ \sqrt{V_t} dW^1_t + dN_t \\
dV_t &= \kappa(\theta - V_t)dt + \sigma \sqrt{V_t} dW^2_t
\end{align*}
where $r$ is the interest rate, $\delta$ is the dividend yield, $V_t$ is the spot volatility, $\theta$ is the long-run volatility, $\kappa$ is the speed of mean-reversion, $\sigma$ is the volatility of volatility, and $W^1_t$ and $W^2_t$ are two correlated Brownian motion with parameter $\rho$. 
The process $N_t$ is a compound Poisson process with intensity $\lambda$ and independent jumps $J$ with
\begin{equation*}
\ln(1+ J) \sim \mathcal{N}\left(\ln(1 + \beta) -\frac{1}{2}\alpha^2, \,  \alpha^2 \right)
\end{equation*}
where the parameters $\alpha$ and $\beta$ determine the distribution of the jumps, and the Poisson process is assumed to be independent of the Brownian motions.

As the characteristic function of the log-price is known, we used the Fast Fourier transform method~\cite{carr1999option} in order to compute option prices efficiently.

\textbf{Calibration details.}
SV models are typically calibrated on options prices since the corresponding implied volatility is not readily available (and would thus require an additional numerical procedure at each iteration).
Note that, for this reason, it is not possible to use SV models as prior models.

We denote here $\pi_{j}$, $\sigma_{j}$, and $\nu_{j}$ the $j$-th option price, implied volatility, and Vega.
Similarly $\hat{\pi}_{j}$ and $\hat{\sigma}_{j}$ denote the model option price and implied volatility.
We calibrate the models by minimizing the Vega-weighted root-mean-square-error (RMSE)
\begin{equation}\label{eq:vegaloss}
\sqrt{\frac{1}{N}\sum_{j=1}^{N} \left(\frac{\pi_{j}-\hat{\pi}_{j}}{\nu_{j}}\right)^2 }
\end{equation}
where $N$ is the number of out-of-the-money options on a particular day.
and where the Vega option Greek is given by
and by\begin{equation*}
\nu = S \e^{-\delta \tau} \phi(d_+)\sqrt{\tau} = K e^{-r \tau} \phi(d_-)\sqrt{\tau}
\end{equation*}
for both Calls and Puts, where $\Phi$ and $\phi$ denotes respectively the standard Gaussian CDF and PDF.

The loss~\eqref{eq:vegaloss} is a computationally efficient approximation for the implied volatility surface RMSE criterion which follows by observing that
\[
\sigma_{j}-\hat{\sigma}_{j} \approx \frac{\pi_{j}-\hat{\pi}_{j}}{\nu_{j}} \quad \text{when} \quad \pi_{j}\approx\hat{\pi}_{j}.
\]


\section{Additional information on model training and data}\label{sec:addit}

\subsection{Training and model parameters} \label{sec:train}

\textbf{Default parameters.}
Unless stated otherwise, the NN will have 4 layers and 40 neurons per layer, the loss penalty values are $\lambda_4=\lambda_5=\lambda_6=10$, and the prior model is the SSVI.
We chose this configuration because it proved to be flexible enough to reproduce many model-based IVS, while always remaining arbitrage-free thanks to the large $\lambda$ penalty value.

\textbf{Parameters initialization.}
We initialize the parameters so that the signal propagated through the layers do not explode or vanish, as motivated in~\cite{glorot2010understanding}.
The parameters $W_{i}$ and $b_i$ are all initialized by Gaussian random variables with mean zero and standard deviation $(\nr{i-1} + \nr{i})^{-1/2}$ where we recall that $\nr{i}$ denotes the output size of layer $i$.

\textbf{Optimization routine.}
The total loss $\Lcal(\theta)$ is minimized with the Adam optimizer \citep{kingma2014adam}. 
As adaptive learning rate and early stopping have shown to significantly improve training \citep{haykin2009neural, hagan1994training, sutskever2013importance, caruana2001overfitting, prechelt1998early, hinton2015distilling}, we follow this approach.
Starting with a learning rate of $0.01$, we let it decrease by a factor $2$ on plateaus of length $500$ epochs when the total loss was not improved by more than $1\%$.
The learning routine stops if $\Lcal(\theta)$ has not improved by $1\%$ over $2'000$ epochs, and restarts using the initial learning rate until $4'000$ total epochs have been reached or until the total loss \eqref{eq:loss} is below $0.25\%$.

Note that the number of epochs is large compared to many deep learning applications, yet the training takes at most few minutes as the training data is also small (at most a few thousand samples at most and only two features).
We are not using minibatch.
In addition, although it takes many epochs for a randomly initialized model to converge to a good solution, we observed that training a model on new data using a previously trained but different model drastically improves the performance so that convergence can be achieved within seconds.
We attribute this fact to the soft constraints which shape the neural network in complex ways that is hardly achievable with random initialization.

\textbf{At the money (ATM) total variance.} 
Because it can be inferred directly from market prices, we consider the ATM total variance term structure $\watm$ a model input as it feeds into the prior models.
Indeed, we expect the prior model to be a first-order approximation of the surface and therefore to at least reproduce ATM values.
This is especially important given that calls/puts close close to ATM are generally the most liquid.

We extract $\watm$ from the market data as follows.
We know that it must be a positive and increasing function of $\tau$ given~\ref{cons:posi} and~\ref{cons:mono}.
For each maturities $\tau$ we collect the total variance $\sigma^2\tau$ values corresponding to the contract closest to $k=0$.
We then use \href{https://www.tensorflow.org/probability/api_docs/python/tfp/math/interp_regular_1d_grid}{\texttt{interp\_regular\_1d\_grid}} from \href{https://www.tensorflow.org/probability}{\texttt{TensorFlow Probability}} if this term-structure is increasing.
Because, for some dates, it was empirically not the case, we use the \href{https://cran.r-project.org/web/packages/scam/scam.pdf}{\texttt{SCAM}} package to fit a spline monotonically increasing spline in $\tau$ with 10 knots and a smoothness penalty being selected by minimizing the generalized cross-validation criterion.
Interpolations and extrapolations of the splines model on a fine grid are then used as constants and fed into 
\href{https://www.tensorflow.org/probability/api_docs/python/tfp/math/interp_regular_1d_grid}{\texttt{interp\_regular\_1d\_grid}} from \href{https://www.tensorflow.org/probability}{\texttt{TensorFlow Probability}}.

As both prior and NN models are trained, we observed that they sometimes compensate each others around the money.
This implies that the ATM prior predictions sometimes deviate from $\watm$.
While this is not an issue, we prefer if $\wpr$ gives the best possible fit and let the NN compensate for its limitations.
Therefore, we added an optional loss function $\Lcal_{\rm atm}$ which encourage the NN model to be close to one for ATM values,
$$
\Lcal_{\rm atm} (\theta) = 
\frac{1}{|\dataatm|} \left( \sum_{(k_i,\tau_i)\in\dataatm} (1 - \wnn(k_i,\tau_i;\theta_1) )^2\right)^{1/2}
$$
for an ATM grid of points $\dataatm$ given by $
\dataatm = \{(0,\tau) \, : \tau \in \Tcal\}$.
We always use a small penalty value of $\lambda_{\rm atm} = 0.1$ for this loss.

\textbf{Feature engineering.}
Feature engineering is not used for the experiments reported in this paper, yet previous results showed that adding some features guided by expert judgment may lead to improved performance for a fixed model capacity.
We observed that including feature variables inversely proportional to the time to maturity allowed to calibrate NN models with fewer layers and neurons for a given accuracy level.
These features would take the form of $k \tau^{-\gamma}$ for $\gamma\in(0,1)$ so that \ref{cons:posi}--\ref{cons:vmat} remain satisfied and that the total variance remains asymptotically linear in $k$.
We conjecture that this is because the implied volatility surface tends to sharply increase with $\lvert k \rvert$ at short horizons while being more flat at longer horizons.

\textbf{Code and hardware.}
The method was implemented using tensorflow \cite{abadi2016tensorflow} and the experiments ran on Tesla K80 GPUs via Amazon Web Services.
The code and data allowing to reproduce the numerical experiment with the synthetic data is provided in the supplementary.
Because the real data was provided by a private provider, it unfortunately cannot be made publicly available.

\subsection{Model based (synthetic) data} \label{sec:datasynt}

To study the properties of our approach in a controlled setting, we create a synthetic dataset using Bates model, see Appendix~\ref{sec:SVJ}.
We use the following grid $\datamkt^{k,\tau}$ of log moneynss and maturities,
$$
\datamkt^{k,\tau} = \big\{(k,\tau) \, : \, k\in\Kcal_0, \,\tau \in \Tcal_0\big\}
$$
with
\begin{align*}
\Kcal_0 &= \big\{\log(x) \, : \, x \in \{0.3, 0.4, 0.6, 0.8, 0.9, 0.95, 0.975, 1, \\
& \qquad\qquad\qquad\qquad  1.025, 1.05, 1.1, 1.2, 1.3, 1.5, 1.75, 2, 2.5, 3 \} \big\}\\
\Tcal_0 & = 
\big\{i/52 \, : \, i \in \{0.5, 1, 2, 3 \} \big\} \cup
\big\{i/12 \, : \, i\in\{1,2,\dots,11,12,18,24\} \big\}  
\end{align*}
or in plain words for $\Tcal_0$: half a week, one, two and three weeks, one to twelve months, eighteen months and two years.

The drift and diffusion parameters in the Bates model are 
$V_0 = 0.10^2$,
$\theta = 0.25^2$,
$\rho = -0.75$,
$\kappa =  0.5$, and
$\sigma = 1$.
and the jump parameters are 
$\lambda = 0.1$,
$\beta = -0.05$, and
$\alpha = 0.15$.
We also set the interest rate and dividend yield to zero.

\subsection{Market data} \label{sec:datamkt}
We use data European option price quotes on the S\&P500 from the CBOE and provided by \emph{OptionMetrics IvyDB US database} through the \emph{Wharton Research Data Services}. Note that the quotes represent the best 15:59EST bid and ask and we use mid-quotes.
We extracted implied volatility values for the periods September-December 2008 and January-April 2018 in multiple steps.

We first estimated the implied risk-free rate and dividend yield values. 
We started by computing the option mid prices by averaging the bid and ask prices, and use it as the reference prices henceforth. 
For each date $t$ and each contract maturity $T$, we use the options around the money ($\pm$ 7.5\%) and estimate coefficient in the linear regression 
$${\rm call}(K,t,T) - {\rm put}(K,t,T) = S_t\,\beta^S_{t,T} + K\beta^K_{t,T}$$
as guided by the put-call parity relation, where $S_t$ is the time-$t$ closing price of the index, also provided by \emph{OptionMetrics}.

We then derive the maturity specific implied risk-free rate $r_{t,T}=-\log(\beta^K_{t,T})/(T-t)$ and dividend yield $\delta_{t,T}=-\log(\beta^S_{t,T})/(T-t)$. 
This allows us to compute the maturity specific log forward moneyness defined by $k = \ln(K/S_t) - (r_{t,T}-\delta_{t,T})(T-t)$ for each option, as well as the implied volatility values. Finally, Brent's method is used to extract the implied volatility on all options in the dataset.

We then apply a set of rules to select a realistic range of option contracts. 
We select only out of the money options, that is call options with $k>0$ and put options with $k<0$. 
We select contracts with time to maturity $(T-t)$ between 2 and 730 days, absolute log forward moyness \(|k|\) less than 1.5, and implied volatility less than 300\%.
Table~\ref{tab:sumstat} provide daily summary statistics.

\begin{table}
\caption{Daily statistics for the implied volatility dataset. \label{tab:sumstat}}
\centering
\begin{tabular}[t]{ccccccc}
\toprule
 &  & Min & Q1 & Median & Q3 & Max\\
\midrule
 & contracts & 3946 & 4405 & 4788 & 4980 & 5291\\
\cmidrule{2-7}
 & maturities & 32 & 33 & 34 & 34 & 35.\\
\cmidrule{2-7}
 & $k$ min & -1.5 & -1.5 & -1.5 & -1.5 & -1.4\\
\cmidrule{2-7}
 & $k$ max & 0.3 & 0.4 & 0.4 & 0.4 & 0.5\\
\cmidrule{2-7}
 & $\sigma_{\rm IV}$ min & 0.0 & 0.1 & 0.1 & 0.1 & 0.1\\
\cmidrule{2-7}
\multirow{-7}{*}{\centering\arraybackslash Jan-Apr/2018} & $\sigma_{\rm IV}$ max & 1.6 & 2.4 & 2.8 & 2.9 & 3.0\\
\cmidrule{1-7}
 & contracts & 513 & 631 & 748 & 790 & 873\\
\cmidrule{2-7}
 & maturities & 12 & 12 & 13 & 13 & 14\\
\cmidrule{2-7}
 & $k$ min & -1.5 & -1.5 & -1.3 & -1.1 & -0.8\\
\cmidrule{2-7}
 & $k$ max & 0.6 & 0.8 & 1.0 & 1.1 & 1.2\\
\cmidrule{2-7}
 & $\sigma_{\rm IV}$ min & 0.2 & 0.2 & 0.2 & 0.3 & 0.3\\
\cmidrule{2-7}
\multirow{-7}{*}{\centering\arraybackslash Sep-Dec/2008} & $\sigma_{\rm IV}$ max & 0.7 & 1.3 & 1.7 & 2.2 & 3.0\\
\bottomrule
\end{tabular}
\end{table}


\section{Additional results}\label{sec:additres}

\subsection{Smoothing behavior} \label{sec:smoothsynt}

\begin{figure}
\centering
\includegraphics[width=\textwidth]{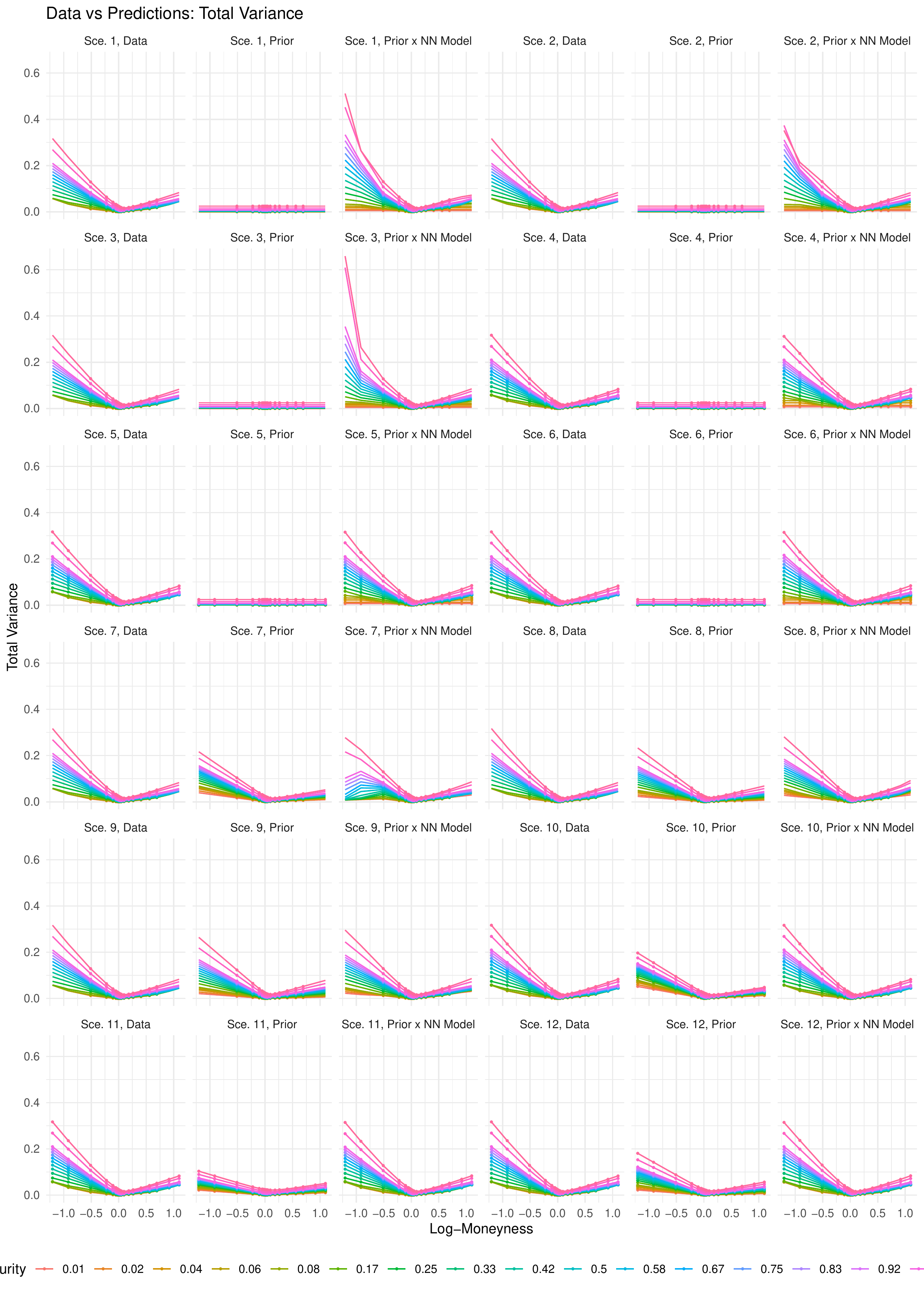}
\caption{\label{fig:smoothsynt:totvar} Synthetic data and trained model total variance predictions.}
\end{figure}

\begin{figure}
\centering
\includegraphics[width=\textwidth]{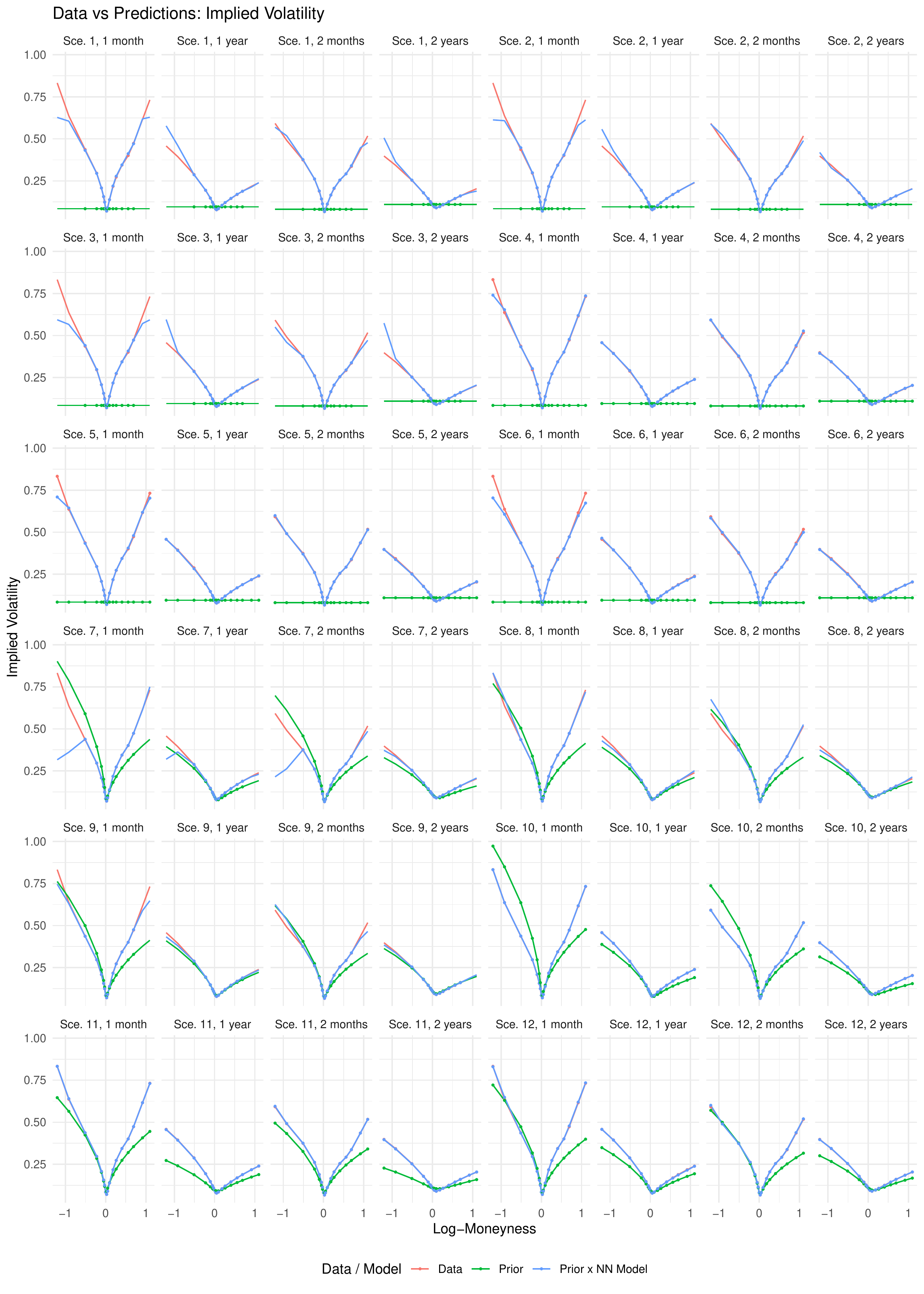}
\caption{\label{fig:smoothsynt:smiles} Synthetic data and trained model implied volatility predictions for selected maturities.}
\end{figure}

\begin{figure}
\centering
\includegraphics[width=\textwidth]{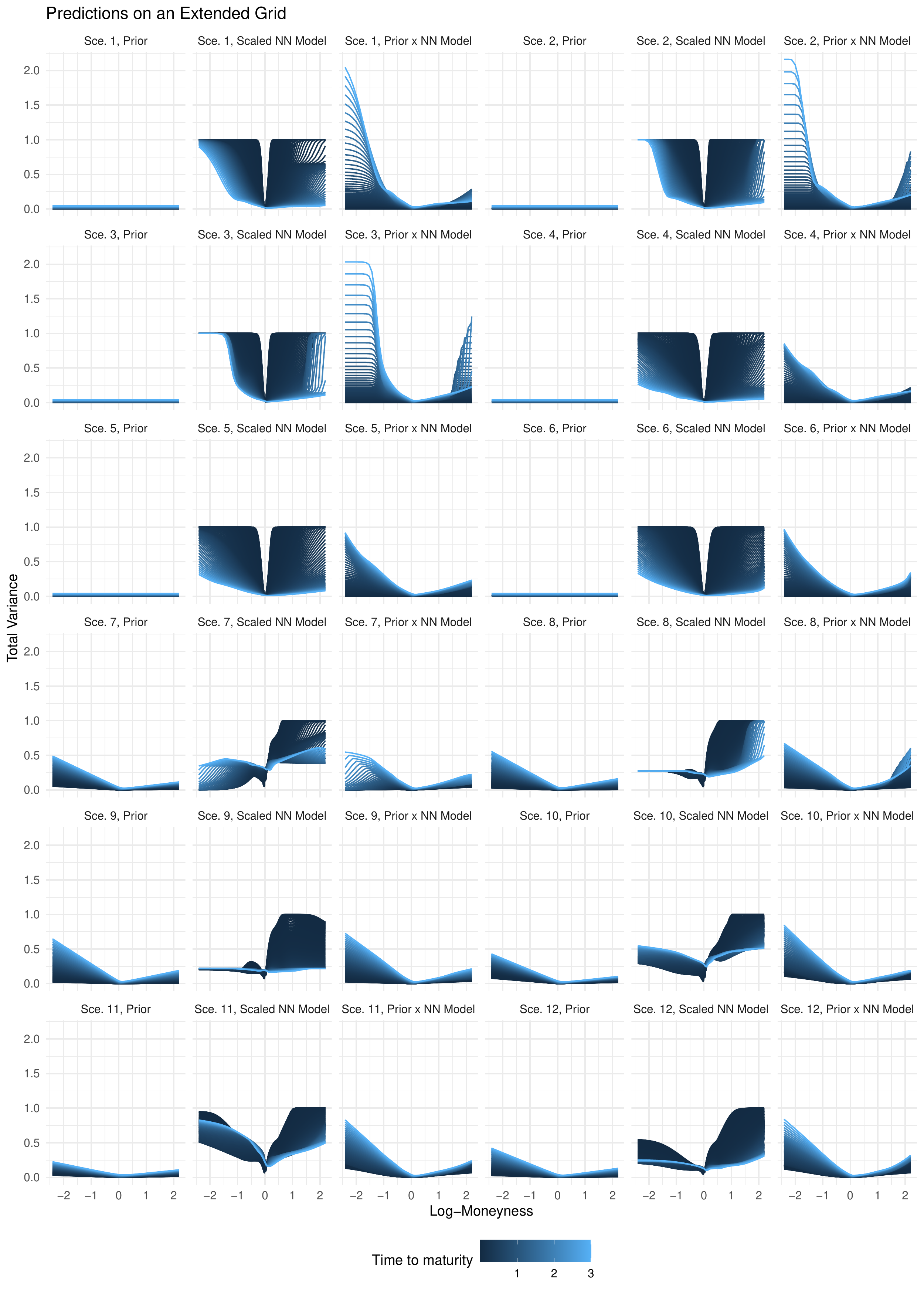}
\caption{\label{fig:smoothsynt:prior} Trained prior and NN predictions for the synthetic data.}
\end{figure}

\Cref{fig:smoothsynt:totvar,fig:smoothsynt:smiles,fig:smoothsynt:prior} display the synthetic data and trained model predictions for the different scenarios summarized in~\Cref{tab:scenarios}.
We observe that the predictions are typically worse with the BS prior than with the SSVI prior.
Also, the predictions outside the observe data region (extrapolations) may exhibit strange behavior when $\lambda=0$ so that the arbitrage opportunities are not controlled.

\begin{table}[ht]
\caption{\label{tab:scenarios} Configurations for the different scenarios, the second column indicates the fraction of near the money data retained for training.}
\centering
\begin{tabular}{lrrrrr}
  \hline
 prior & $k$ prop. & $\lambda_4$ & $\lambda_5$ & $\lambda_6$ & scenario \\ 
  \hline
BS & 0.8 & 0 & 0 & 0 & 1 \\ 
BS & 0.8 &  1 & 1 & 1 & 2 \\ 
BS & 0.8 & 10 & 10 & 10 & 3 \\ 
BS & 1 & 0 & 0 & 0 & 4 \\ 
BS & 1 & 1 & 1 & 1 & 5 \\ 
BS & 1 & 10 & 10 & 10 & 6 \\ 
SSVI & 0.8 & 0 & 0 & 0 & 7 \\ 
SSVI & 0.8 & 1 & 1 & 1 & 8 \\ 
SSVI & 0.8 & 10 & 10 & 10 & 9 \\ 
SSVI & 1 & 0 & 0 & 0 & 10 \\ 
SSVI & 1 & 1 & 1 & 1 & 11 \\ 
SSVI & 1 & 10 & 10 & 10 & 12 \\
   \hline
\end{tabular}
\end{table}


\subsection{Losses and convergence} \label{sec:convergence}

\begin{figure}
\centering
\includegraphics[width=\textwidth]{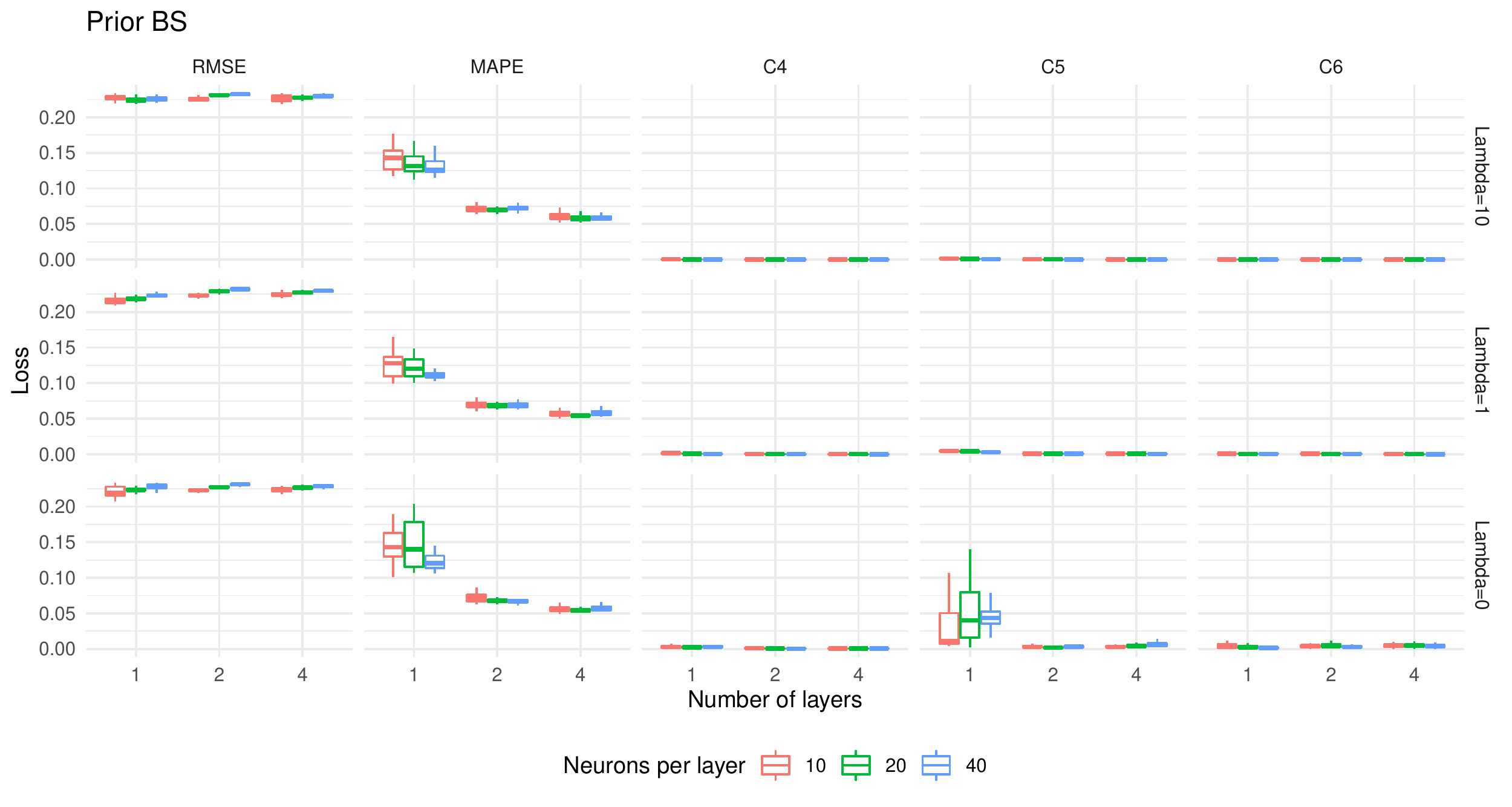}
\caption{\label{fig:convbs} Losses for different number of layers, neurons per layer, and penalty value}
\end{figure}

\Cref{fig:convbs} displays the different losses for the BS prior using 50 random seeds and 5'000 epochs for each configuration.
It highlights that the NN model does not succeed to compensate for the poor choice of prior model given the data.

\subsection{Comparison with Dugas et al.'s NN model} \label{sec:dugasetal}

\begin{figure}
\centering
\includegraphics[width=\textwidth]{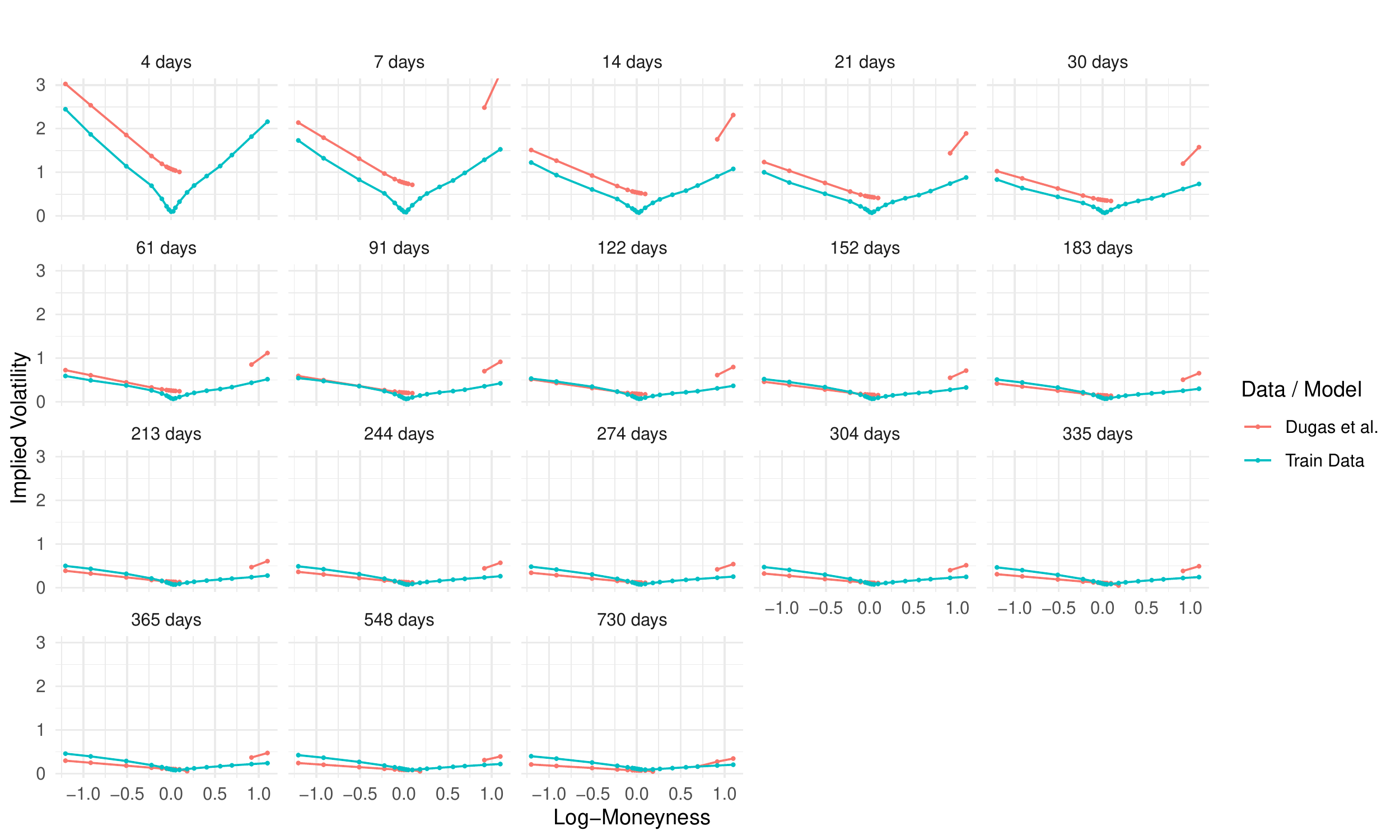}
\caption{IV surface for the Dugas et al.\ NN model trained on the synthetic data. \label{fig:dugas_iv}}
\end{figure}

\begin{figure}
\centering
\includegraphics[width=\textwidth]{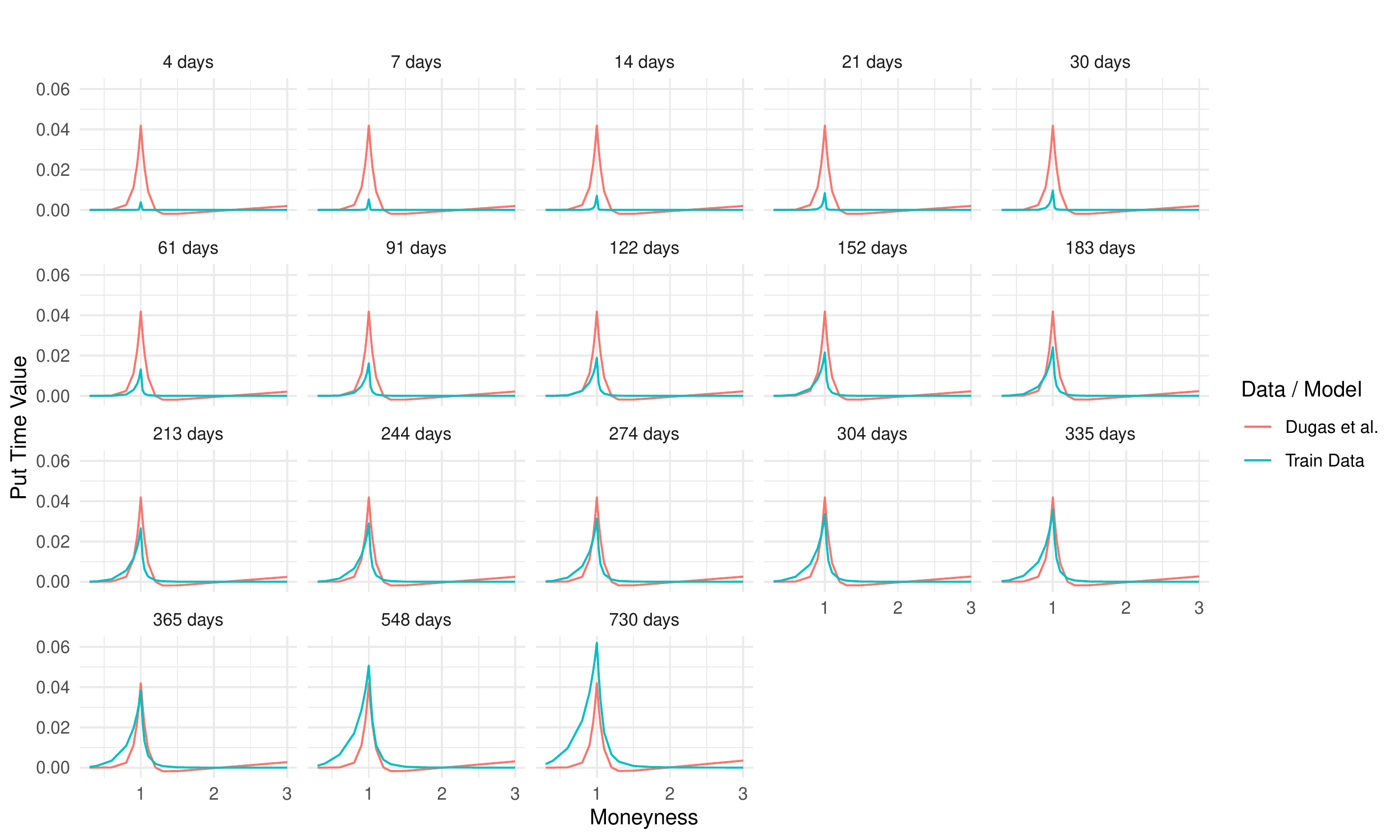}
\caption{Put option time value for the Dugas et al.\ NN model trained on the synthetic data. \label{fig:dugas_put}}
\end{figure}

In \cite{dugas2001incorporating} the authors proposed a one-layer neural network model to approximate the call, or put, price surface which is arbitrage-free by construction.
We trained this model with 1'000 neurons for 50'000 epochs on the put price surface from the Bates synthetic data, see~\Cref{sec:datasynt}.
\Cref{fig:dugas_iv} displays the train and fitted implied volatility surfaces.
We observe large IV errors for the Dugas et al.\ model, indeed some IV values are no even within the range $(0.01\%, \, 10'000\%)$ and are thus not displayed.
This is because small price errors for far out of the money options translate into large implied volatility errors.
In addition an option whose time value is smaller than zero is erroneous and cannot be mapped to an IV value, recall that the time value is defined as the difference between the option price and its intrinsic value ($(K - S_0)^+$ here).
\Cref{fig:dugas_put} highlights this issue for the put prices predicted by fitted the Dugas et al.\ model.

\subsection{Impact of arbitrage opportunities on the losses} \label{sec:numpexpao}

\begin{figure}
\centering
\includegraphics[width=\textwidth]{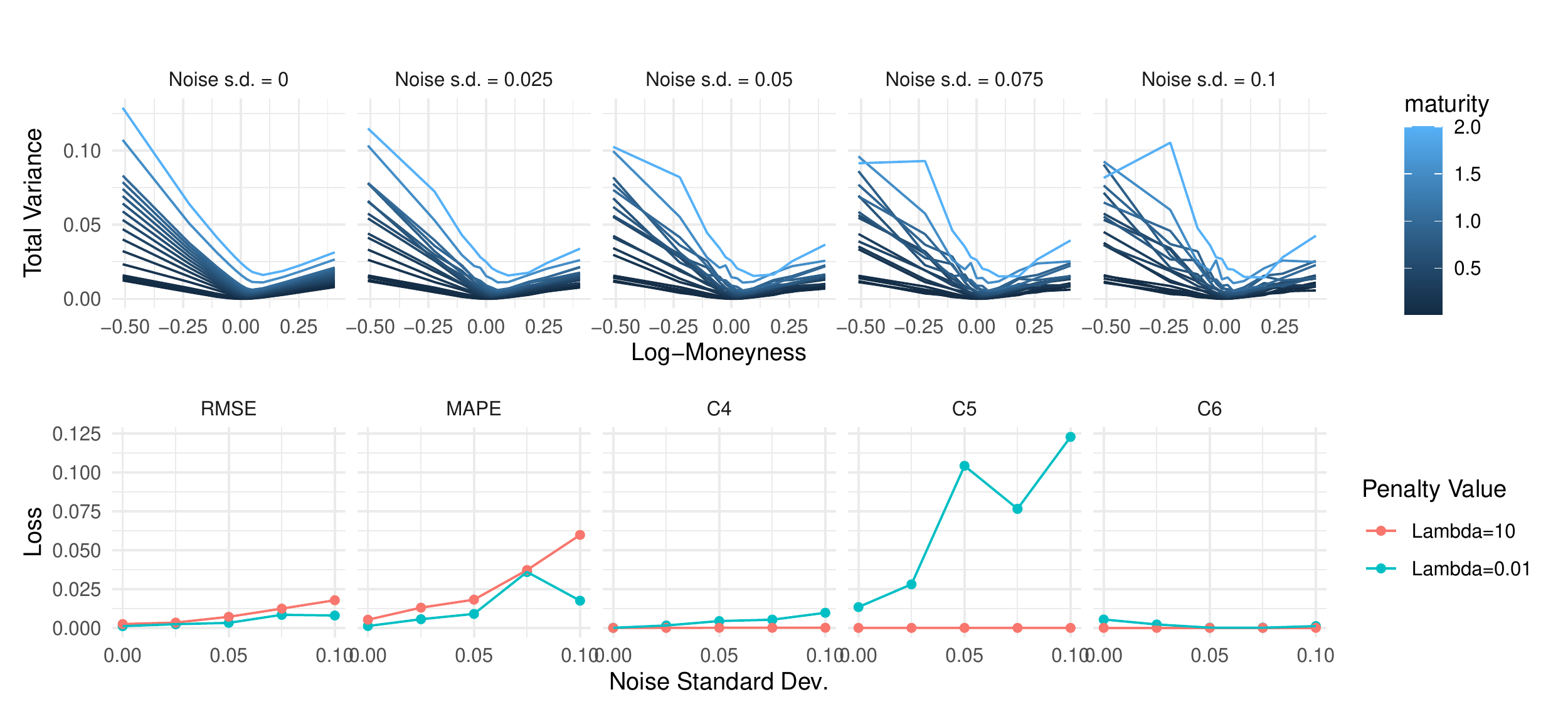}
\caption{Noise impact on the IVS shape and trained NN losses. \label{fig:numpexpao}}
\end{figure}

We illustrate the impact of erroneous IVS data on the different losses by adding random Gaussian noise with volatility parameter $\eta$ to the synthetic log-IV data.
That is, if $\sigma$ is an IV value then we multiply it by $\exp(\epsilon)$ where $\epsilon\sim\mathcal{N}(0,\eta)$.
The first row of \Cref{fig:numpexpao} displays the perturbed total variance surfaces, and the second row different losses for the trained NN model with penalty value $\lambda\in\{0.01,\, 10\}$.
We observe that a larger noise level translates into larger prediction errors for the $\lambda=10$ model as the more severe arbitrage opportunities are smoothed away. 
The prediction errors for $\lambda=0.01$ model also increases with $\eta$, although at a lower rate, because its capacity remains fixed while the surface complexity further increases.

\subsection{Smoothing real data} \label{sec:smoothmkt}

\begin{figure}
\centering
\includegraphics[width=\textwidth]{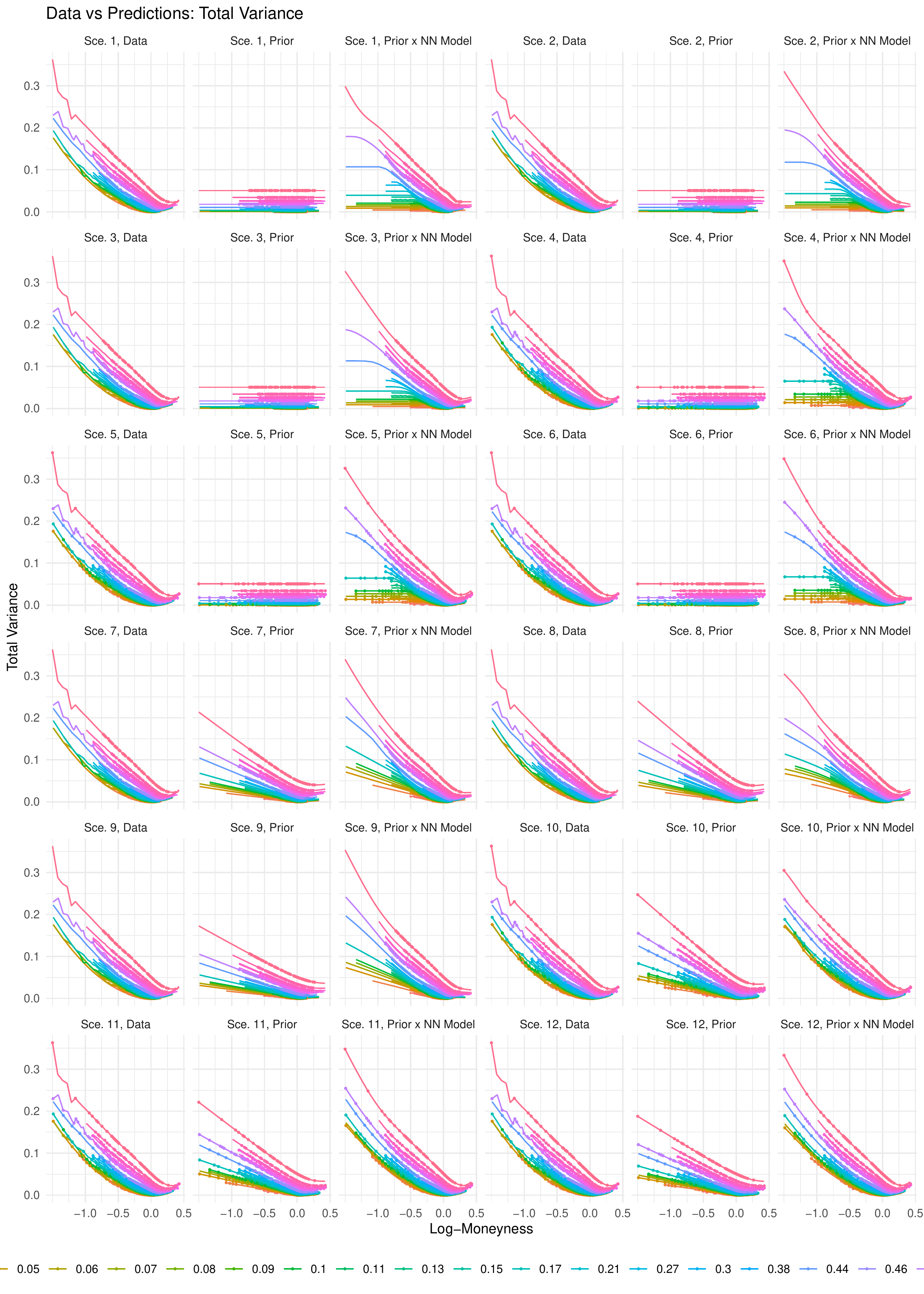}
\caption{\label{fig:smoothmkt:totvar} Market data and trained model total variance predictions.}
\end{figure}

\begin{figure}
\centering
\includegraphics[width=\textwidth]{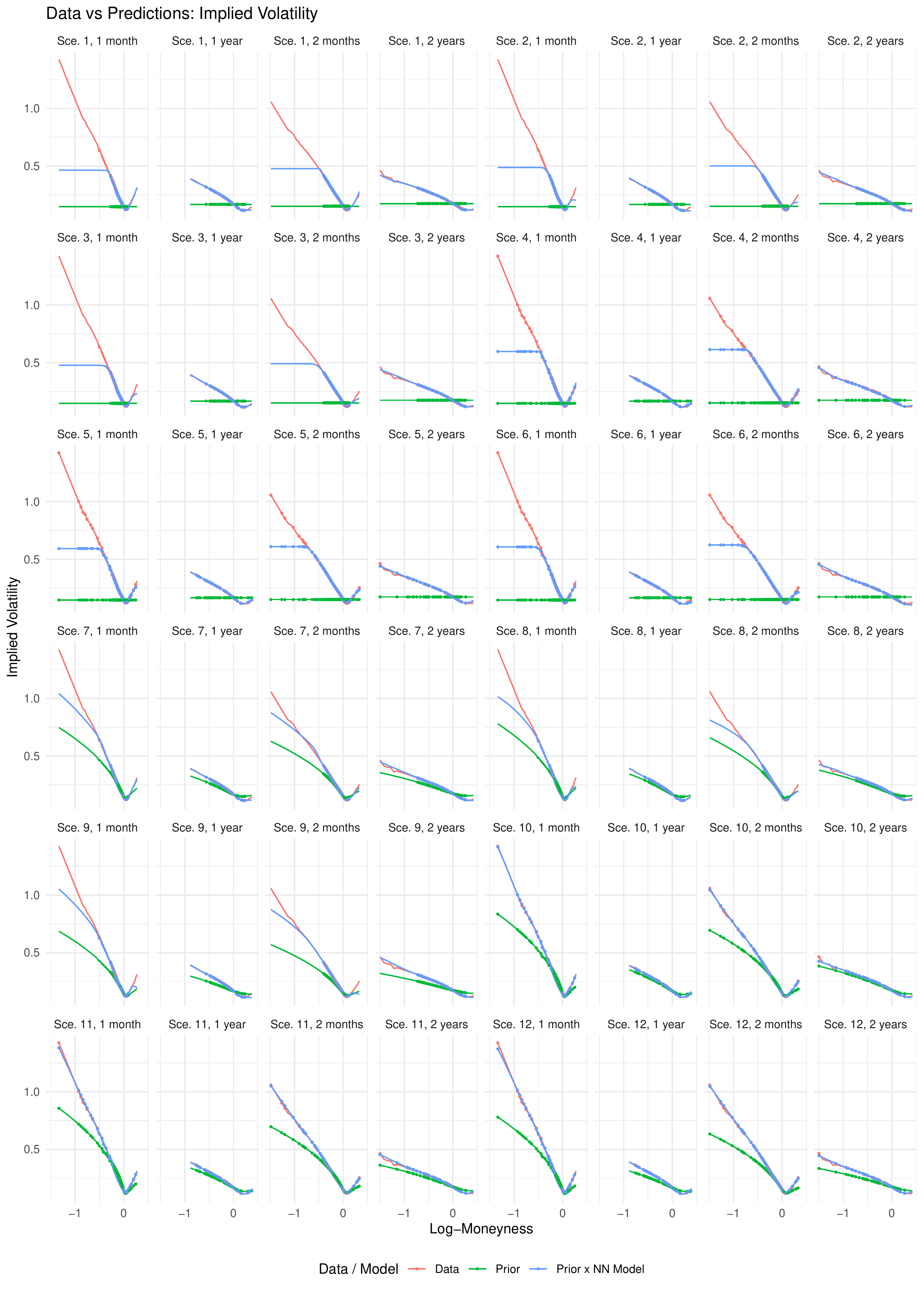}
\caption{\label{fig:smoothmkt:smiles} Market data and trained model implied volatility predictions for selected maturities.}
\end{figure}

\begin{figure}
\centering
\includegraphics[width=\textwidth]{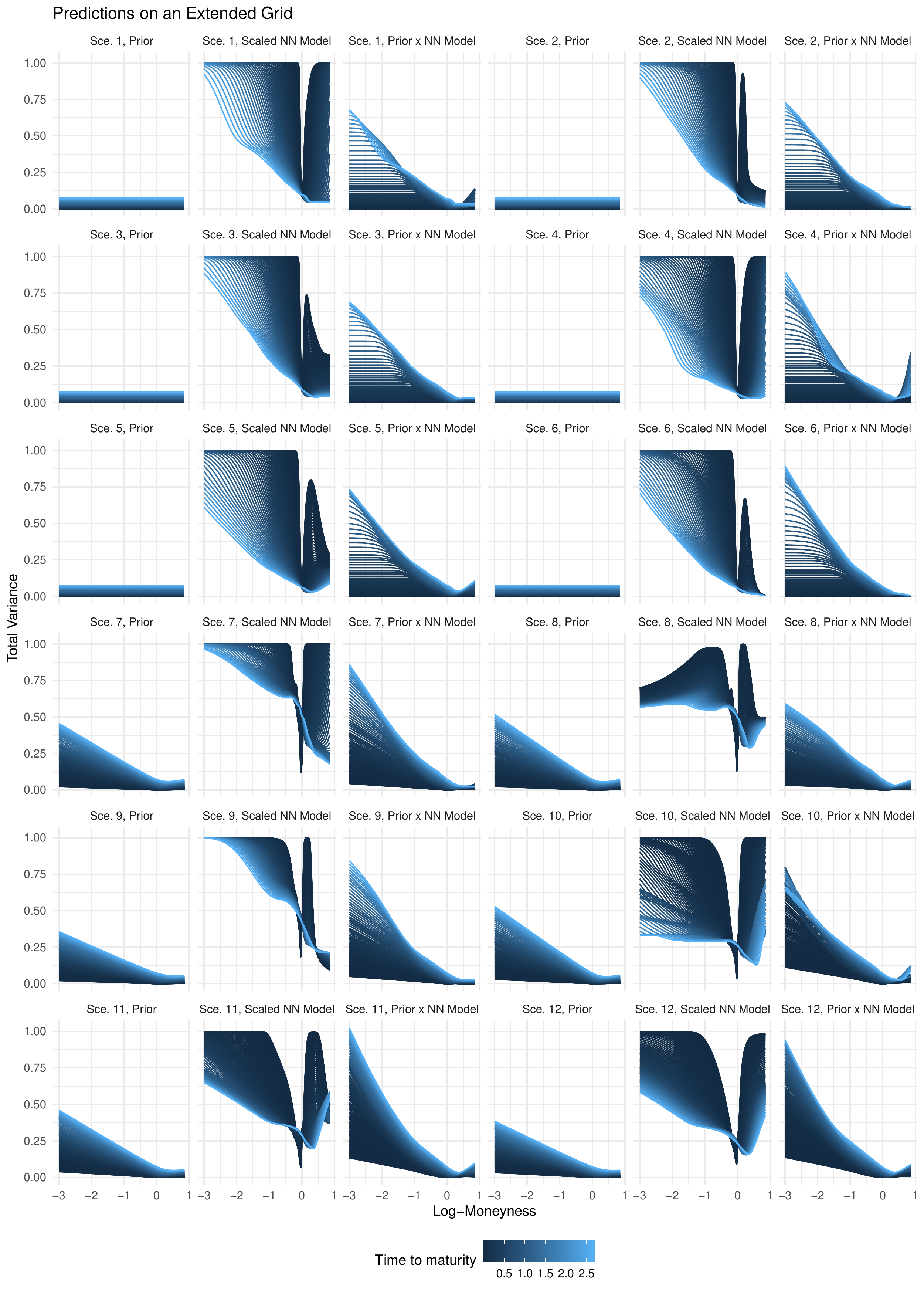}
\caption{\label{fig:smoothmkt:prior} Trained prior and NN predictions for the market data.}
\end{figure}

\Cref{fig:smoothmkt:totvar,fig:smoothmkt:smiles,fig:smoothmkt:prior} display the market data and trained model predictions for the different scenarios summarized in~\Cref{tab:scenarios}.
The conclusions are similar as for the synthetic data.
We observe that the choice of prior has a notable impact on the NN model, and that the NN modle is always significantly more accurate than its corresponding prior.


\subsection{Backtests} \label{sec:backtests}

\Cref{tab:bktBS}~displays the backtesting results with the Black-Scholes prior.
\Cref{tab:bktbench}~displays the backtesting results for the benchmark models (Bates and SSVI).

\begin{table}
\caption{\label{tab:bktBS}Backtesting results for the BS prior (quantiles in \%, Jan-Apr/2018 - Sep-Dec/2008)}
\centering
\resizebox{\textwidth}{!}{
\begin{tabular}[t]{cccccccccccccc}
\toprule
\multicolumn{1}{c}{ } & \multicolumn{1}{c}{ } & \multicolumn{6}{c}{Interpolation} & \multicolumn{6}{c}{Extrapolation} \\
\cmidrule(l{3pt}r{3pt}){3-8} \cmidrule(l{3pt}r{3pt}){9-14}
\multicolumn{1}{c}{ } & \multicolumn{1}{c}{ } & \multicolumn{3}{c}{Train} & \multicolumn{3}{c}{Test} & \multicolumn{3}{c}{Train} & \multicolumn{3}{c}{Test} \\
\cmidrule(l{3pt}r{3pt}){3-5} \cmidrule(l{3pt}r{3pt}){6-8} \cmidrule(l{3pt}r{3pt}){9-11} \cmidrule(l{3pt}r{3pt}){12-14}
Loss & $\lambda$ & $q_{05}$ & $q_{50}$ & $q_{95}$ & $q_{05}$ & $q_{50}$ & $q_{95}$ & $q_{05}$ & $q_{50}$ & $q_{95}$ & $q_{05}$ & $q_{50}$ & $q_{95}$\\
\midrule
 & 10 & 0.9 / 0.7 & 3.6 / 2.1 & 14.3 / 13.2 & 1.1 / 1.0 & 3.8 / 2.7 & 14.7 / 14.8 & 0.2 / 0.3 & 0.4 / 1.2 & 4.2 / 5.1 & 5.4 / 7.4 & 10.8 / 16.6 & 21.1 / 31.2\\
\cmidrule{2-14}
 & 1 & 0.5 / 0.7 & 3.0 / 18.3 & 14.4 / 39.1 & 0.6 / 1.4 & 2.9 / 18.9 & 14.9 / 44.4 & 0.2 / 0.3 & 0.3 / 1.1 & 4.4 / 3.3 & 6.0 / 9.0 & 11.5 / 16.6 & 21.6 / 29.4\\
\cmidrule{2-14}
\multirow{-3}{*}{\centering\arraybackslash RMSE} & 0 & 0.4 / 0.3 & 2.8 / 0.8 & 14.4 / 7.6 & 0.6 / 1.8 & 3.0 / 5.0 & 14.3 / 25.2 & 9.5 / 0.3 & 12.7 / 0.8 & 15.4 / 3.1 & 24.6 / 15.2 & 29.7 / 34.2 & 35.7 / 69.5\\
\cmidrule{1-14}
 & 10 & 0.9 / 1.1 & 1.2 / 1.8 & 4.3 / 7.7 & 0.9 / 1.3 & 1.3 / 2.0 & 4.4 / 6.9 & 0.5 / 0.5 & 0.8 / 1.0 & 2.3 / 4.5 & 2.5 / 3.8 & 3.7 / 5.5 & 8.3 / 12.0\\
\cmidrule{2-14}
 & 1 & 0.6 / 1.2 & 0.9 / 22.2 & 4.4 / 29.2 & 0.7 / 1.5 & 0.9 / 23.0 & 4.2 / 29.6 & 0.4 / 0.5 & 0.7 / 1.0 & 2.4 / 2.7 & 2.5 / 3.7 & 3.9 / 5.4 & 8.8 / 10.2\\
\cmidrule{2-14}
\multirow{-3}{*}{\centering\arraybackslash MAPE} & 0 & 0.6 / 0.5 & 0.8 / 0.8 & 4.0 / 1.5 & 0.6 / 1.4 & 0.9 / 2.2 & 4.0 / 4.3 & 24.2 / 0.4 & 31.3 / 0.8 & 40.4 / 1.7 & 33.3 / 7.1 & 40.9 / 18.5 & 50.7 / 34.4\\
\cmidrule{1-14}
 & 10 & 0.0 / 0.0 & 0.0 / 0.0 & 0.1 / 0.3 & 0.0 / 0.0 & 0.0 / 0.0 & 0.1 / 0.3 & 0.0 / 0.0 & 0.0 / 0.0 & 0.0 / 0.0 & 0.0 / 0.0 & 0.3 / 0.0 & 1.4 / 0.9\\
\cmidrule{2-14}
 & 1 & 0.0 / 0.0 & 0.1 / 0.0 & 0.3 / 2.3 & 0.0 / 0.0 & 0.1 / 0.0 & 0.3 / 3.7 & 0.0 / 0.0 & 0.0 / 0.0 & 0.1 / 0.3 & 0.0 / 0.0 & 0.7 / 0.1 & 1.9 / 13.2\\
\cmidrule{2-14}
\multirow{-3}{*}{\centering\arraybackslash C4} & 0 & 0.8 / 12.1 & 11.1 / 96.4 & 75.2 / 99+ & 0.6 / 12.6 & 11.3 / 99+ & 39.4 / 99+ & 0.0 / 7.7 & 0.0 / 37.9 & 0.0 / 99+ & 0.0 / 10.0 & 0.0 / 27.8 & 0.0 / 99+\\
\cmidrule{1-14}
 & 10 & 0.0 / 0.0 & 0.0 / 0.0 & 0.0 / 0.0 & 0.0 / 0.0 & 0.0 / 0.0 & 0.0 / 0.1 & 0.0 / 0.0 & 0.0 / 0.0 & 0.0 / 0.0 & 0.0 / 0.0 & 3.5 / 0.0 & 15.9 / 0.8\\
\cmidrule{2-14}
 & 1 & 0.0 / 0.0 & 0.0 / 0.0 & 0.1 / 7.5 & 0.0 / 0.0 & 0.0 / 0.0 & 0.1 / 99+ & 0.0 / 0.0 & 0.0 / 0.0 & 0.0 / 0.3 & 0.2 / 0.0 & 3.7 / 0.6 & 16.5 / 8.4\\
\cmidrule{2-14}
\multirow{-3}{*}{\centering\arraybackslash C5} & 0 & 1.2 / 91.8 & 20.0 / 99+ & 99+ / 99+ & 1.2 / 99+ & 21.5 / 99+ & 99+ / 99+ & 0.0 / 99+ & 0.0 / 99+ & 0.0 / 99+ & 0.0 / 99+ & 0.0 / 99+ & 0.0 / 99+\\
\cmidrule{1-14}
 & 10 & 0.0 / 0.0 & 0.0 / 0.0 & 0.0 / 0.1 & 0.0 / 0.0 & 0.0 / 0.0 & 0.0 / 0.8 & 0.0 / 0.0 & 0.0 / 0.0 & 0.0 / 0.0 & 0.0 / 0.0 & 0.0 / 0.0 & 0.0 / 0.0\\
\cmidrule{2-14}
 & 1 & 0.0 / 0.0 & 0.0 / 0.0 & 0.0 / 0.2 & 0.0 / 0.0 & 0.0 / 0.0 & 0.1 / 33.5 & 0.0 / 0.0 & 0.0 / 0.0 & 0.0 / 0.0 & 0.0 / 0.0 & 0.0 / 0.0 & 0.0 / 4.1\\
\cmidrule{2-14}
\multirow{-3}{*}{\centering\arraybackslash C6} & 0 & 2.3 / 7.0 & 37.1 / 99+ & 99+ / 99+ & 2.0 / 10.7 & 33.9 / 99+ & 99+ / 99+ & 0.0 / 0.0 & 0.0 / 69.9 & 0.0 / 99+ & 0.0 / 0.0 & 0.0 / 0.0 & 0.0 / 99+\\
\bottomrule
\end{tabular}
}
\end{table}

\begin{table}
\caption{\label{tab:bktbench}Backtesting results for the benchmarks (quantiles in \%, Jan-Apr/2018 - Sep-Dec/2008)}
\centering
\resizebox{\textwidth}{!}{
\begin{tabular}[t]{llllllllllllll}
\toprule
\multicolumn{1}{c}{ } & \multicolumn{1}{c}{ } & \multicolumn{6}{c}{Interpolation} & \multicolumn{6}{c}{Extrapolation} \\
\cmidrule(l{3pt}r{3pt}){3-8} \cmidrule(l{3pt}r{3pt}){9-14}
\multicolumn{1}{c}{ } & \multicolumn{1}{c}{ } & \multicolumn{3}{c}{Train} & \multicolumn{3}{c}{Test} & \multicolumn{3}{c}{Train} & \multicolumn{3}{c}{Test} \\
\cmidrule(l{3pt}r{3pt}){3-5} \cmidrule(l{3pt}r{3pt}){6-8} \cmidrule(l{3pt}r{3pt}){9-11} \cmidrule(l{3pt}r{3pt}){12-14}
Benchmark & Loss & $q_{05}$ & $\mu$ & $q_{95}$ & $q_{05}$ & $\mu$ & $q_{95}$ & $q_{05}$ & $\mu$ & $q_{95}$ & $q_{05}$ & $\mu$ & $q_{95}$\\
\midrule
 & RMSE & 1.6 / 2.0 & 2.9 / 5.1 & 5.3 / 10.8 & 1.7 / 2.1 & 2.8 / 5.2 & 5.0 / 9.7 & 1.0 / 1.2 & 1.6 / 3.1 & 4.3 / 8.1 & 2.4 / 2.5 & 4.6 / 7.0 & 8.7 / 14.4\\
\cmidrule{2-14}
\multirow[t]{-2}{*}{\raggedright\arraybackslash BATES} & MAPE & 3.7 / 3.5 & 6.1 / 5.9 & 11.8 / 11.4 & 3.6 / 3.5 & 6.1 / 6.1 & 11.8 / 11.5 & 3.2 / 2.4 & 5.6 / 4.5 & 20.6 / 12.6 & 4.1 / 4.0 & 6.4 / 7.2 & 18.1 / 15.6\\
\cmidrule{1-14}
 & RMSE & 4.0 / 3.1 & 6.7 / 7.3 & 10.8 / 15.0 & 4.0 / 2.9 & 7.0 / 7.3 & 10.6 / 15.7 & 1.5 / 1.8 & 2.2 / 4.6 & 4.0 / 11.3 & 6.6 / 5.0 & 11.2 / 10.8 & 16.2 / 21.4\\
\cmidrule{2-14}
\multirow[t]{-2}{*}{\raggedright\arraybackslash SSVI} & MAPE & 5.1 / 4.3 & 7.2 / 8.6 & 9.7 / 13.8 & 5.0 / 4.1 & 7.2 / 8.7 & 9.6 / 13.3 & 3.2 / 2.0 & 4.6 / 6.5 & 7.2 / 11.9 & 6.5 / 5.9 & 8.8 / 9.9 & 11.2 / 14.1\\
\bottomrule
\end{tabular}
}
\end{table}

\subsection{Risk-neutral density and local volatility} \label{sec:pdflocvol}

\begin{figure}
\centering
\includegraphics[width=0.75\textwidth]{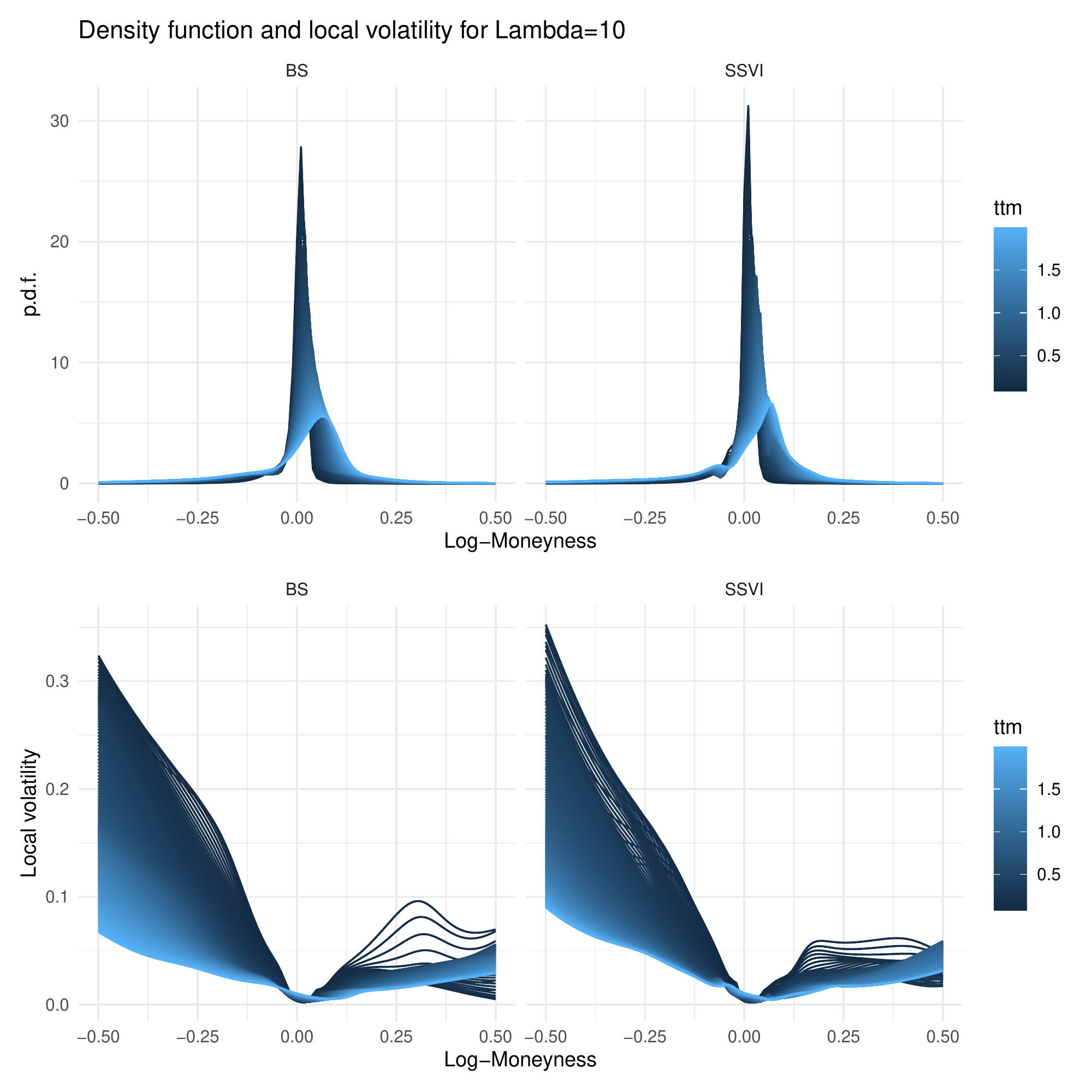}
\caption{\label{fig:lvpdf} Density function and local volatility for scenarios 6 and 12.}
\end{figure}

We apply the approaches described in~\ref{sec:fineng} for a model fitted on the synthetic data of~\ref{sec:datasynt}.
Figure~\ref{fig:lvpdf} displays the price density (first row) and the local volatility (second row) with respect to the log moneyness and for different horizons.

\subsection{W-shaped smile} \label{sec:wshape}

\begin{figure}
\centering
\includegraphics[width=0.75\textwidth]{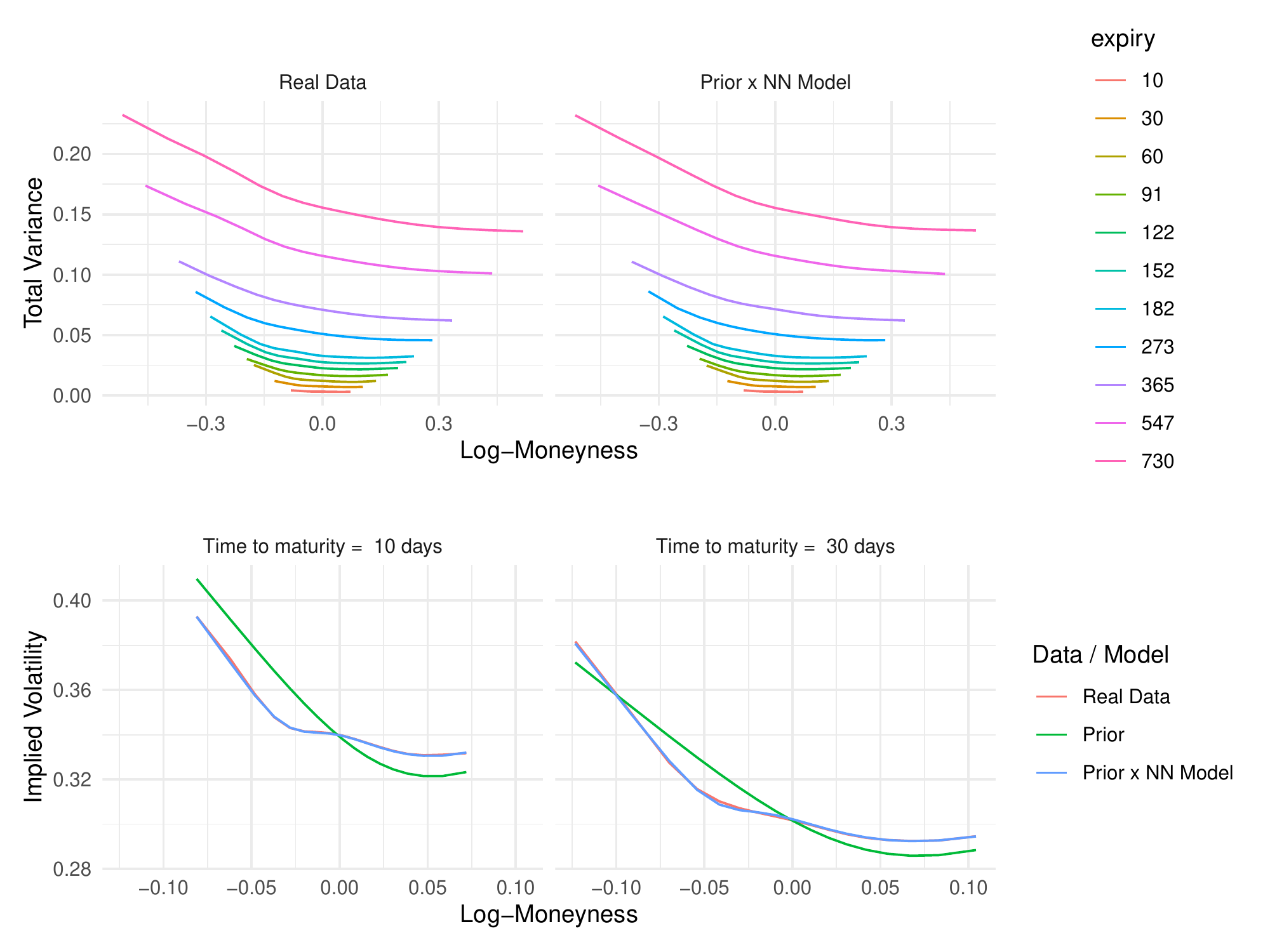}
\caption{W-shaped smiles for the options on the Apple stock on July 21st 2015. \label{fig:wfit}}
\end{figure}

We extracted from \emph{OptionMetrics} IVS database the implied volatility values for call options on Apple on the July 21st 2015.
In this database the option prices have already been mapped into IV values for standardized strikes and maturities, that is we have 11 maturities and 17 strikes per maturities.
The particularity of this selected surface is that it exhibits W-shaped IV smiles.
This sometimes happens before scheduled news releases such as company earnings.
These are challenging surfaces to fit for classical models.
We calibrate our model with the SSVI prior on the entire surface.
\Cref{fig:wfit} displays the fitted total variance surface, as well as the fitted smiles for the first two maturities for which we observe that the SSVI prior fails to capture the W-shaped smile whereas the NN model succeeds.

\subsection{Spread weighted errors} \label{sec:spreadw}

\begin{figure}
\centering
\includegraphics[width=0.75\textwidth]{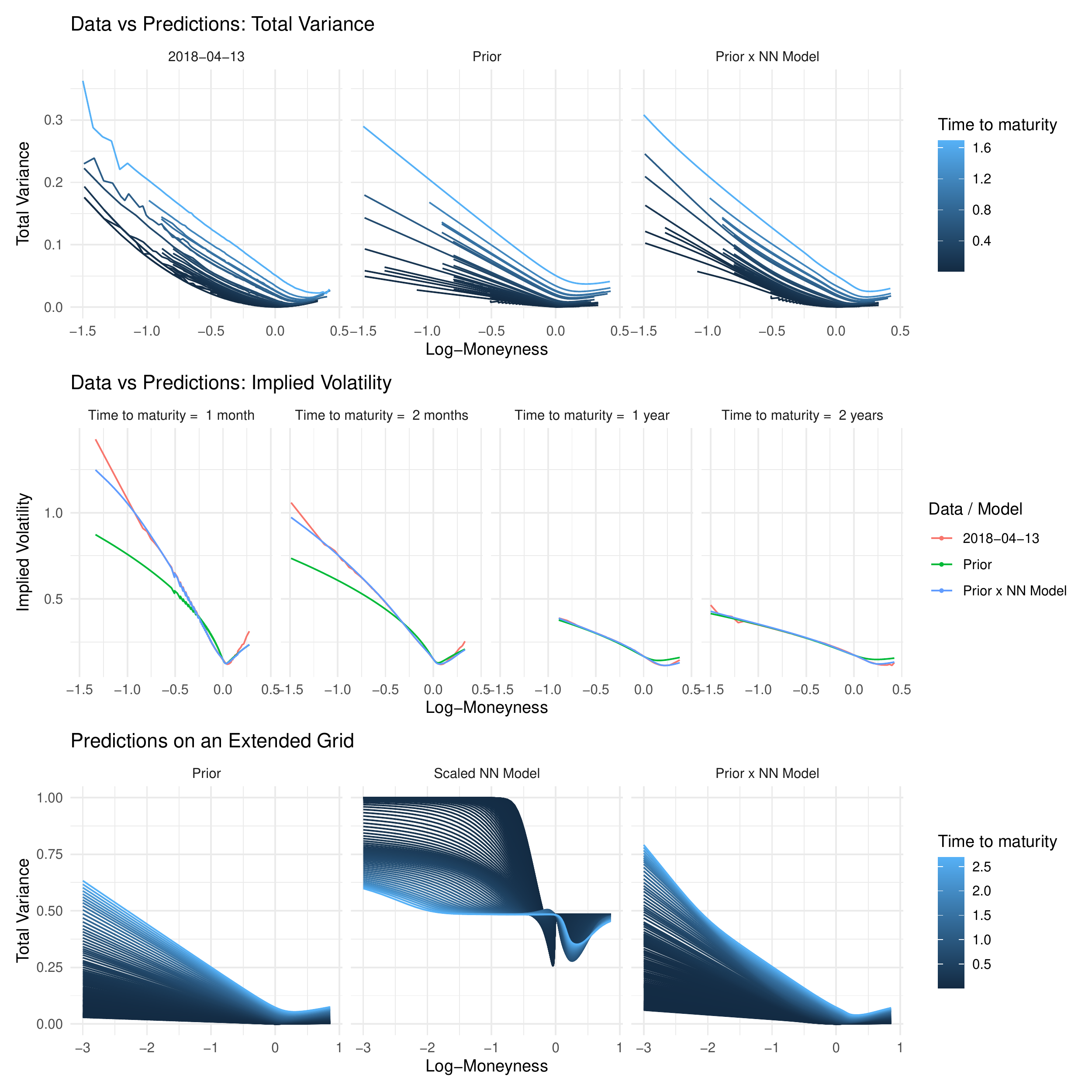}
\caption{Market data and trained model predictions for a specific configuration (scenario 12) and trained on spread weighted errors. \label{fig:spreadw}}
\end{figure}

We train the NN model with a different loss function, 
\[
\Lcal_0(\theta) = 1/|\datamkt| \sum_{(\sigma_i,k_i, \tau_i)\in\datamkt} \frac{|\sigma_i - \ivth(k_i, \tau_i)|}{1 + \sigma^{\rm ask}_i - \sigma^{\rm bid}_i}
\]
where $\sigma^{\rm bid}_i, \sigma^{\rm ask}_i$ are the implied volatility corresponding to bid and ask option prices, and $\sigma_i$ corresponds to the option mid-price.
\Cref{fig:spreadw} displays the results for scenario 12.
We observe that the predictions for far out of the money options are not as accurate as with the default loss function.
This is not surprising as far out of the money options typically have large IV spread values.

\subsection{Computational time} \label{sec:time}

\Cref{tab:times} reports the summary statistics for the NN model training times.
Note that the implementation producing the results reported in this paper calls Python TensorFlow from R, hence experiences overhead.
In addition, the code itself was not written to be efficient, but rather for exploratory flexibility. 
In our experience, a simple TensorFlow implementation in Python would typically runs at least 2-4 times faster.

\begin{table}

\caption{\label{tab:times}Computation times (in minutes)}
\centering
\begin{tabular}[t]{cccccc}
\toprule
Prior & Period & Min & Mean & Sd & Max\\
\midrule
 & Sep-Dec/2008 & 1.0 & 1.3 & 0.6 & 4.7\\
\cmidrule{2-6}
\multirow{-2}{*}{\centering\arraybackslash BS} & Jan-Apr/2018 & 0.9 & 1.4 & 0.7 & 6.6\\
\cmidrule{1-6}
 & Sep-Dec/2008 & 1.8 & 2.1 & 0.9 & 8.0\\
\cmidrule{2-6}
\multirow{-2}{*}{\centering\arraybackslash SSVI} & Jan-Apr/2018 & 1.5 & 2.2 & 1.0 & 8.0\\
\bottomrule
\end{tabular}
\end{table}

\section{Discussion}\label{sec:discussion}

\subsection{Extensions}

The presented approach focuses on a single stock and observations at a single time because this already represents a challenging and relevant practical problem.
Still, we discuss how one could approach the modeling of multiple stocks, and the IVS at multiple time points.

\textbf{Multi-asset.}
It is straightforward to extend the current approach to create a model for multiple IVSs.
Indeed, one may consider a model taking as input an input categorical variable that will the volatility surface.
This approach would allow to transfer reuse and transfer features learned from on surface to another, which could be highly beneficial in situations where few observations are available for IVS.
This advantage typically comes at the cost of building and training a deeper and larger NN for a joint model than for individual models.
Note that, additional no-arbitrage constraints may have to be imposed in the specific case of implied volatility surfaces for currency pairs.

\textbf{Multi-day.}
Modeling the IVS at multiple time points could be useful to transfer information forward from the past, which may be allows to build a more resilient image of the IVS a given time point.
This may be achieved in different ways.
For example, the parameters $\theta$ of the previous model could be used as initial values for the model to be newly fitted.
Alternatively, one may think of propagating forward in time a latent factor as performed in recurrent neural networks.

Another application could the construction of a generative model for future IVSs which could be used for risk-management, for example.
Such a generative model would be required to simulate entire IVSs given present and past values.
Our approach could be used in combination with a generative module to guarantee that the fake IVSs are sensible.
\end{appendices}

\end{document}